\documentclass[aps,prd,twocolumn,superscriptaddress,nofootinbib,floatfix,preprintnumbers]{revtex4-2}

\usepackage{amsmath}
\usepackage{bm}
\usepackage{graphicx}
\usepackage[svgnames]{xcolor}
\usepackage{subcaption}
\usepackage{caption}
\usepackage{aas_macros}
\usepackage{soul}

\newcommand{\revold}[1]{}

\def\E{{\cal E}}

\newcommand{\edit}[2]{\textcolor{red}{\st{#1}}\textcolor{blue}{#2}}  

\graphicspath{{./}{plots/}}

\begin{document}

\preprint{MIT-CTP/6029}

\title{%
Effects of magnetically driven shocks on nucleosynthesis and kilonovae\\ from neutron star mergers
}

\author{Yuan Feng}
\email{yfeng2@caltech.edu}
\affiliation{TAPIR, Mailcode 350-17, California Institute of Technology, Pasadena, CA 91125, USA}

\author{Oleg Korobkin}
\affiliation{Center for Theoretical Astrophysics, Los Alamos National Laboratory, Los Alamos, NM 87545 USA}

\author{Elias R. Most}
\email{emost@caltech.edu}
\affiliation{TAPIR, Mailcode 350-17, California Institute of Technology, Pasadena, CA 91125, USA}
\affiliation{Walter Burke Institute for Theoretical Physics, California Institute of Technology, Pasadena, CA 91125, USA}

\author{Ananda F. Smith}
\affiliation{TAPIR, Mailcode 350-17, California Institute of Technology, Pasadena, CA 91125, USA}
\affiliation{Center for Theoretical Physics – a Leinweber Institute,
Massachusetts Institute of Technology, Cambridge, MA 02139, USA}

\author{Christopher J. Fontes}
\affiliation{Center for Theoretical Astrophysics, Los Alamos National Laboratory, Los Alamos, NM 87545 USA}
\affiliation{Computational Physics Division, Los Alamos National Laboratory, P.O. Box 1663, Los Alamos, New Mexico 87545, USA}


\begin{abstract}
Neutron-star mergers can launch mildly relativistic to moderately relativistic outflows whose interaction with the ejecta can reshape kilonova emission.
We parametrically study magnetically powered outbursts from long-lived merger remnants, such as flare-like eruptions and collapse-driven shocks, and quantify their impact on ejecta dynamics, composition, and observables.
Using two-dimensional special-relativistic magnetohydrodynamic simulations, we follow magnetized blast waves injected into expanding merger ejecta for early- and late-launch scenarios across a range of shock strengths.
We then post-process Lagrangian tracer histories with the nuclear reaction network \texttt{WinNet} and the radiative-transfer code \texttt{SuperNu} with realistic opacities, to connect shock heating directly to nucleosynthesis and kilonova light curves.
We find that sufficiently strong shocks can reheat portions of the ejecta to nuclear statistical equilibrium, increase the electron fraction in the shocked material, and deposit entropy, leading to systematic changes in $r$-process yields.
These thermodynamic and compositional changes can leave observable imprints on kilonova emission---especially in color evolution and late-time light-curve behavior---indicating that magnetically driven remnant variability can potentially contribute to kilonova diversity.
\end{abstract}

\maketitle

\section{Introduction}\label{sec:intro}

The binary neutron-star merger event GW170817 has firmly established compact-binary coalescences as multi-messenger sources \citep{LIGOScientific:2017vwq, LIGOScientific:2017ync}. In addition to the gravitational-wave signal, GW170817 showed a prompt short-duration gamma-ray burst and a rapidly evolving kilonova counterpart in the ultraviolet/optical/near-infrared bands \citep{LIGOScientific:2017zic, Savchenko:2017ffs, Cowperthwaite:2017dyu, Nicholl:2017ahq, Chornock:2017sdf, Tanvir:2017pws, Villar:2017wcc, Kasliwal:2017ngb, 2017Natur.551...75S, 2017Natur.551...67P}, followed by long-lived non-thermal X-ray and radio emission requiring mildly relativistic to relativistic outflows \citep{Margutti:2017cjl, Hallinan:2017woc, Mooley:2017enz, Mooley:2018qfh, Hajela:2019mjy}.

Kilonova emission is understood as thermal reprocessing of radioactive heating following nucleosynthesis in neutron-rich merger ejecta, produced during and after the collision \citep{Li:1998bw, Metzger:2019zeh}.
On the modeling side, a robust qualitative picture has emerged in which blue and red kilonova components encode different electron fractions, velocities, and opacities \citep{Nicholl:2017ahq, Tanvir:2017pws, Villar:2017wcc, Metzger:2019zeh}, with multidimensional radiative-transfer and geometry-focused studies highlighting strong viewing-angle-dependent structure \citep{Collins:2022ocl, Shingles:2023kua, Collins:2023btn}.
Substantial work has also been put into analysis and inference pipelines for future detections \citep{Villar:2017wcc, Cowperthwaite:2017dyu, Metzger:2019zeh, Ricigliano:2023svx}.

While substantial work has been devoted to understanding uncertainties in these models, e.g., the atomic opacities used or uncertainties related to different mass ejection channels from the neutron-star remnant \cite{Radice:2020ddv, Ciolfi:2017uak} or the accretion disk \cite{Radice:2020ddv, Most:2021ytn}, some aspects remain less well explored, including intrinsic strong source variability, which could provide a one-off or repeated injection of additional energy into the ejecta.
This matters because reheating of the ejecta could alter its nuclear composition and in turn the heating rate powering the kilonova; see, e.g., Ref. \cite{Villar:2017wcc, Troja:2018ruz} for questions surrounding sphericity or Ref. \cite{2024MNRAS.533.2303G} on the emergence of strongly anisotropic shocks in the ejecta.

\captionsetup[table]{format=plain,justification=raggedright,singlelinecheck=false}

\begin{table*}[t]
  \centering
  \renewcommand{\arraystretch}{1.15}
  \resizebox{\textwidth}{!}{%
  \begin{tabular}{l c c c c c c c c c c}
    \hline
    simulations & $ \sqrt{4 \pi} c^2 A_0/\sqrt{G}$ & $\gamma_{\max}$ & $r_{c}/r_{0}$ & $r_{e}/r_{0}$ & $v_{c}/c$ & $v_{e}/c$ & $\rho_{c}\,(\mathrm{g\,cm^{-3}})$ & $\rho_{e}\,(\mathrm{g\,cm^{-3}})$ & time after merger & $E_{\rm mag}/M_{\rm ej} c^2$ \\
    \hline
    early\_A & $2\times 10^{-3}$ & 1.4 & 67.72   & 677.22  & $0.2$ & $0.5$ & $3.23\times10^{11}$ & $1.79\times10^{2}$   & 40\,ms  & $4.24\times10^{-2}$ \\
    early\_B & $3\times 10^{-3}$ & 2.0 & 67.72   & 677.22  & $0.2$ & $0.5$ & $3.23\times10^{11}$ & $1.79\times10^{2}$   & 40\,ms  & $9.55\times10^{-2}$ \\
    early\_C & $4\times 10^{-3}$ & 3.3 & 67.72   & 677.22  & $0.2$ & $0.5$ & $3.23\times10^{11}$ & $1.79\times10^{2}$   & 40\,ms  & $1.70\times10^{-1}$ \\
    early\_D & $5\times 10^{-3}$ & 4.4 & 67.72   & 677.22  & $0.2$ & $0.5$ & $3.23\times10^{11}$ & $1.79\times10^{2}$   & 40\,ms  & $2.65\times10^{-1}$ \\
    late\_A  & $2\times 10^{-3}$ & 1.1  & 2097.96 & 5752.85 & $0.2$ & $0.5$ & $2.81\times10^{11}$ & $2.92\times10^{-2}$ & 540\,ms & $2.80\times10^{-2}$ \\
    late\_B  & $3\times 10^{-3}$ & 1.2  & 2097.96 & 5752.85 & $0.2$ & $0.5$ & $2.81\times10^{11}$ & $2.92\times10^{-2}$ & 540\,ms & $6.31\times10^{-2}$ \\
    late\_C  & $4\times 10^{-3}$ & 1.3  & 2097.96 & 5752.85 & $0.2$ & $0.5$ & $2.81\times10^{11}$ & $2.92\times10^{-2}$ & 540\,ms & $1.12\times10^{-1}$ \\
    late\_D  & $5\times 10^{-3}$ & 1.5  & 2097.96 & 5752.85 & $0.2$ & $0.5$ & $2.81\times10^{11}$ & $2.92\times10^{-2}$ & 540\,ms & $1.75\times10^{-1}$ \\
    \hline
  \end{tabular}
  }
  \caption{Simulation parameters for the early- and late-launch models. Here, $A_0$ is the dimensionless strength parameter of the initial vector potential
  $\gamma_{\rm max}$ is the maximum Lorentz factor reached in the simulation right after the shock passed and the temperature dropped below 1\,MeV, $r_c$ is the slow-fast ejecta boundary, and $r_e$ is the outer cloud radius. {The radii are normalized to $r_0$, which is the inner edge of the ejecta cloud}. The quantities $v_c$ and $v_e$ denote the velocities at the slow-fast ejecta edge and the outer cloud edge, respectively, while $\rho_c$ is the core rest-mass density and $\rho_e$ is the rest-mass density at the outer cloud edge. We also list the time after merger at which the blast wave is launched. The final column gives the ratio of magnetic energy, $E_{\rm mag}$, stored inside $r<r_0$ to ejecta rest-mass, $M_{\rm ej}$.}
  \label{tab:sim_params}
\end{table*}

Additional sources of energy injection into the ejecta can roughly be split into two broad scenarios:
impulsive energy injection, either through one-off ejections caused by the collapse of the remnant or a similar cataclysmic event \cite{Lehner:2011aa,Most:2024qgc}, or through flares or other outbursts from the remnant \cite{Most:2023sft} (see also Ref. \cite{Beloborodov:2013kpa} for a scenario in the context of white dwarf mergers), and continued energy injection from a long-lived magnetar \cite{2014MNRAS.439.3916M, Lasky:2013yaa, DallOsso:2014hpa}.
While each of these scenarios can leave distinct observational signatures (e.g., from the presence of pulsar wind termination shocks \cite{2014MNRAS.439.3916M}), determining which additional channels are present depends on the total mass (a long-lived proto-magnetar-like remnant or black hole formation) and on additional microphysics in the remnant. \\

In this work, we focus on one particular class of models, namely magnetically driven flares \cite{Most:2023sft, 2024PhRvD.109f4061N, 2025PhRvD.111j3043J} and magnetospheric shocks \cite{Beloborodov:2022pvn, Most:2024qgc}, recently demonstrated to accompany the collapse of very long-lived merger remnants.
While these so-called monster shocks \cite{Beloborodov:2022pvn, Most:2024qgc, 2025ApJ...982L..54K} are capable of producing X-ray and radio emission \cite{Most:2024qgc, Beloborodov:2022pvn}, in the dense environment of the merger it is most likely that they will be fully absorbed \cite{Beloborodov:2020ylo, Most:2024qgc}, and instead will reheat the ejecta.\\

While full source modeling of the flare and shock launching scenario requires multi-physics numerical relativity simulations of neutron star mergers (see e.g. Refs. \cite{Baiotti:2016qnr, Radice:2020ddv} for recent reviews, and Refs. \cite{Kiuchi:2015sga, Mosta:2020hlh, Ciolfi:2017uak, Giacomazzo:2013uua, Giacomazzo:2014qba} for simulations with magnetic field dynamics included), the large-scale dynamics can be efficiently modeled in dimensionally reduced scenarios, as is common for ejecta-heating simulations~\cite{Metzger:2019zeh} and jet simulations targeting ejecta interactions~\cite{Gottlieb:2017pju, Xie:2018vya}.
See also Ref. \citep{Nathanail:2018dcl, DuPont:2024sbz} for related scenarios to the one discussed here.

In order to understand the general trend, as well as potential constraints on any magnetically driven shock-like outflows, we use two-dimensional magnetohydrodynamic (MHD) simulations to parametrically investigate the scenario of a magnetically launched outflow interacting with an ejecta cloud.
We then systematically post-process ejecta fluid elements with the nucleosynthesis code \texttt{WinNet}\cite{,Reichert:2023xqy}, showing that sufficiently strong shocks reheat the ejecta, alter the resulting nucleosynthesis, and change the heating rate of the kilonova, which we also calculate. This, in turn, leads to a different radioactive heating profile and a different kilonova signature. We calculate the kilonova using the radiative-transfer code \texttt{SuperNu} with detailed atomic binned opacities~\cite{2013ApJS..209...36W, Wollaeger:2017ahm, 2021ascl.soft03019W, 2020MNRAS.493.4143F, 2023MNRAS.519.2862F,nist_opac}.\\

Our work is structured as follows. In Sec. \ref{sec:methods} we detail our setup, before presenting results on the shock evolution and the kilonova in Sec. \ref{sec:results}. We then conclude in Sec. \ref{sec:conclusions}.


\section{Methods}\label{sec:methods}

In this work, we numerically model the interaction of a strongly magnetized blast wave with a post-merger neutron-star ejecta cloud. We do so parametrically, by modeling different ejecta profiles and different launch times of the blast wave. As a matter of convention, in this work we consider ``early'' and ``late'' launch scenarios, corresponding to flare \cite{Most:2023sft} and (monster) shock cases \cite{Beloborodov:2022pvn, Most:2024qgc, 2025ApJ...982L..54K} that naturally occur at different times after merger.

Specifically, in the early-launch models we trigger the shock at $t \simeq 40~\mathrm{ms}$, when the outflow is still comparatively dense and compact, and its radial structure remains close to the immediate post-merger state.
In the late-launch models, by contrast, the shock is launched at $t \simeq 540~\mathrm{ms}$, after continued expansion has made the background ejecta more dilute and extended.
For each launch scenario, we systematically vary the injected energy by adjusting the initial magnetic field strength at the wave launching point, which in turn sets the Lorentz factor attained by the blast wave at later times. This model suite then allows us to isolate how (i) the shock dynamics and mixing depend on the evolving ejecta background, (ii) the associated heating and acceleration reshape the thermodynamic histories of the outflow, (iii) the resulting thermodynamic processing imprints on the final nuclear composition, and (iv) these changes propagate to the predicted kilonova emission.

Because our primary goal is to isolate shock-driven heating and the ensuing hydrodynamic response of an approximately axisymmetric blast, especially at late times, we evolve the system in two spatial dimensions. The length scales of interest are far larger than the gravitational radius of the central remnant, and we restrict to times well before fallback becomes important \cite{Radice:2020ddv}, so that the shock propagation and its interaction with the expanding ejecta can be modeled with special-relativistic magnetohydrodynamics (SRMHD).

\begin{figure}[t]
    \centering
    \includegraphics[width=\columnwidth]{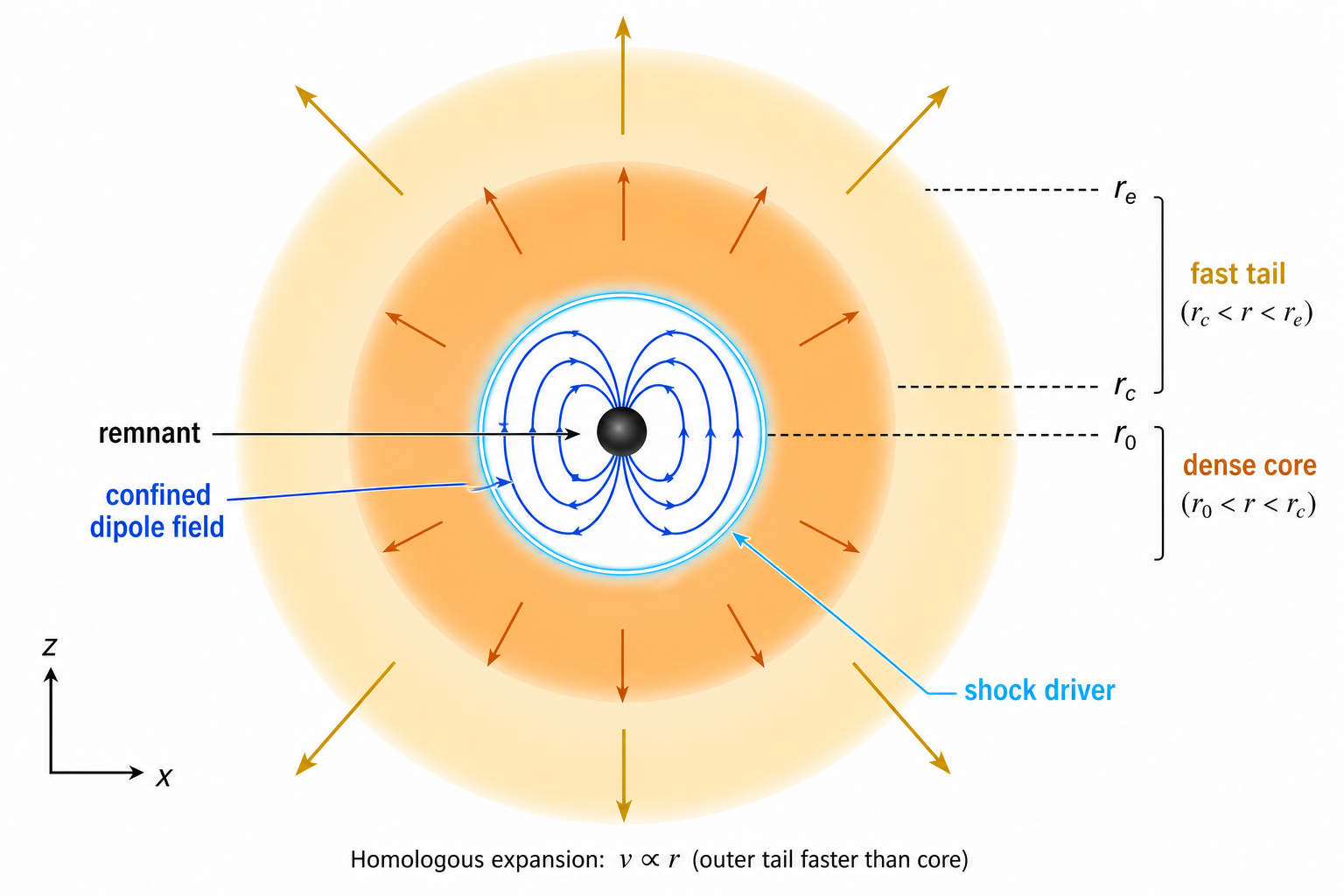}
    \caption{
        Schematic of the initial setup. A dipolar magnetic field is initialized only inside the truncation radius $r_0$, while the ejecta cloud begins outside this radius. The cloud is decomposed into a dense inner core, $r_0<r<r_c$, and a diffuse fast tail, $r_c<r<r_e$, with homologous radial expansion.}
    \label{fig:initial_setup_cartoon}
\end{figure}

The geometry of the initial setup used below is summarized schematically in Fig. \ref{fig:initial_setup_cartoon}.

\begin{figure*}[t]
    \centering
    \includegraphics[width=\textwidth]{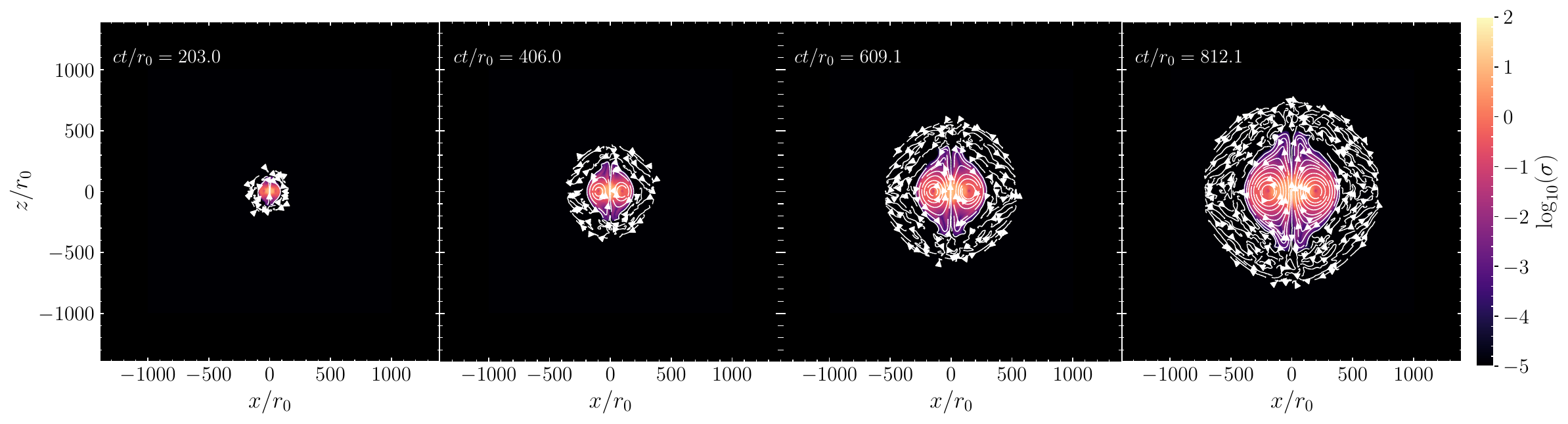}
    \caption{
        Initial launch of a blast wave in the early-launch scenario with maximum Lorentz factor $\gamma_{\max}=2.0$. Shown in color is the magnetization parameter $\sigma$, at different times $t$ after launch, relative to the inner cloud radius crossing time $r_0/c$. White streamlines denote the in-plane magnetic field.}
    \label{fig:sigma}
\end{figure*}

Our focus is on the downstream shock--ejecta interaction rather than the detailed trigger of the magnetically powered disturbance. We therefore omit the computationally intensive modeling of the collapse or flare engine itself and instead initiate the shock via an imposed, localized pressure perturbation embedded in a dipolar magnetic-field configuration, initialized via a magnetic vector potential \cite{Shibata:2011fj}
\begin{align}
    A_{\phi} = -\frac{\sqrt{4 \pi} c^2}{\sqrt{G}} A_{0} \varpi_0\frac{(\varpi/\varpi_0)^2}{\left(\left(\frac{\varpi}{\varpi_0}\right)^2+ 1\right)^{3/2}}\,,
\end{align}
for $\varpi < \varpi_{0}$, and $A_{\phi}=0$ otherwise, where $\varpi=\sqrt{x^{2}+y^{2}}$ is the cylindrical radius, $A_0$ is the dimensionless strength parameter, and $\varpi_0$ is a regularization scale.
The magnetosphere is initialized with a magnetic moment aligned with the $+z$ axis and a uniform magnetization $\sigma \equiv B^{2}/(4 \pi \rho c^2)=40$, where $B$ is the magnetic-field strength and $\rho_{m}$ is the magnetospheric mass density.
Unlike purely magnetospheric shock cases, the important parameter for this work is the Lorentz factor reached inside the ejecta cloud, which depends not just on the initial Lorentz factor of the magnetospheric shock (which is set by $\sigma$), but also on the ratio of magnetospheric energy in the shock compared to the (rest-mass) energy of the ejecta cloud.
Since for our work the details of the launching mechanism are not relevant, we choose to simply keep $\sigma$ fixed but adjust the overall field strength until a target Lorentz factor is reached.

Accordingly, our initial conditions comprise two components: (i) an expanding ejecta cloud and (ii) a magnetically powered shock driver that launches a strong MHD wave into the ejecta.

For the latter, we construct a dipole magnetic field, which we truncate at radius $r_0$, beyond which we begin the ejecta cloud. We seed an initial pressure perturbation $\Delta p$, thereby launching the shock. While artificial in principle, this clean separation allows us to launch the blast wave magnetospherically in a controlled way. Since the shock moves close to the speed of light, we neglect further mass ejection after the shock has been launched.
Our cloud setup follows established models in the literature \citep{Gottlieb:2017pju, Murguia-Berthier:2017kkn}. In detail, this cloud consists of two distinct regions: an inner core extending from $r_0$ to $r_c$, and an outer diffuse \textit{fast tail} that spans from $r_c$ to $r_e$ \citep{Gottlieb:2017pju, Murguia-Berthier:2017kkn}.
We then adopt density and velocity profiles consistent with three-dimensional numerical relativity simulations of binary neutron star mergers \citep{Radice:2020ddv, Baiotti:2016qnr, Kiuchi:2014hja}. Specifically, we adopt a two-zone model, consisting of a fast tail representing dynamical ejecta \citep{Radice:2020ddv}, and a slow inner core, representing secular ejecta \citep{Metzger:2019zeh, Ciolfi:2017uak}. For simplicity we further choose $\varpi_0=r_0$.

The inner core exhibits a mildly oblate density distribution described by the power-law \citep{Gottlieb:2017pju, Murguia-Berthier:2017kkn}:
\begin{equation}
\rho(r) = \rho_c \left(\frac{r}{r_0}\right)^{-n}\left(0.25 + \sin(\theta)^3\right),
\end{equation}
where $\theta \in [0, \pi]$ denotes the polar angle measured from the $z$-axis, and we set $n \approx 3$. The core expands radially with a homologous velocity profile, i.e., one that increases linearly with radius \citep{Gottlieb:2017pju, Murguia-Berthier:2017kkn}:
\begin{equation}
v(r) = \left(\frac{r}{r_c}\right)v_c,
\end{equation}
with $v_c$ approximately an order of magnitude below the speed of light $c$. The outer fast tail region features a similar initial velocity gradient transitioning from $v_c$ at $r_c$ to a higher velocity $v_e > v_c$ at $r_e$, governed by:
\begin{equation}
v(r) = v_c\left(1 - \frac{r - r_c}{r_e - r_c}\right) + v_e\left(\frac{r - r_c}{r_e - r_c}\right).
\end{equation}
The density profile of the fast tail declines steeply with velocity according to:
\begin{equation}
\rho(r) = \rho_e \left(\frac{v(r)}{v_e}\right)^{-m},
\end{equation}
with $n < m = 4$. This behavior is in accordance with numerical relativity simulations of the merger process \citep{Radice:2020ddv, Kiuchi:2014hja}. Finally, the ambient background region beyond $r_e$ is set to a small atmospheric value $\rho_{\rm atmo}=6.17\times10^{-5}~\mathrm{g\,cm^{-3}}$. All simulation parameters explored in this work are listed in Tab. \ref{tab:sim_params}.

In order to integrate the equations of SRMHD numerically, we adopt the \texttt{AthenaK} code framework \citep{2026ApJS..283...27S}.
\texttt{AthenaK} uses standard high-resolution shock-capturing ingredients for relativistic MHD---including piecewise-parabolic reconstruction \cite{1984JCoPh..54..174C}, the HLLE Riemann solver \cite{Einfeldt1988}, and second-order Runge-Kutta timestepping \cite{Butcher2008}.
We use the tabulated finite-temperature SFHo equation of state \cite{Steiner:2012rk} from the CompOSE database, consistent with modern merger calculations \citep{Palenzuela:2015dqa, Radice:2020ddv}.
Our simulation domain consists of a Cartesian static-mesh-refinement grid.
For the early-launch models, the computational domain extends to $\pm 1,\!002\,r_0$ and employs 7 equally spaced refinement levels in total.
For the late-launch models, the domain extends to $\pm 7,\!991\,r_0$ and employs 9 equally spaced refinement levels in total.
In both cases, we use 2048 grid points per direction on each refinement level.
The smallest refined box has a width of $15.62\,r_0$, which yields a finest grid spacing $\Delta x \simeq 7.65\times10^{-3}\,r_0$. Once the initial shock has been launched and left the central launch region, we de-refine the central part of the domain as it does not participate in any of the relevant dynamics.

\begin{figure*}[t]
    \centering

    \begin{subfigure}{\textwidth}
        \centering
        \includegraphics[width=\textwidth]{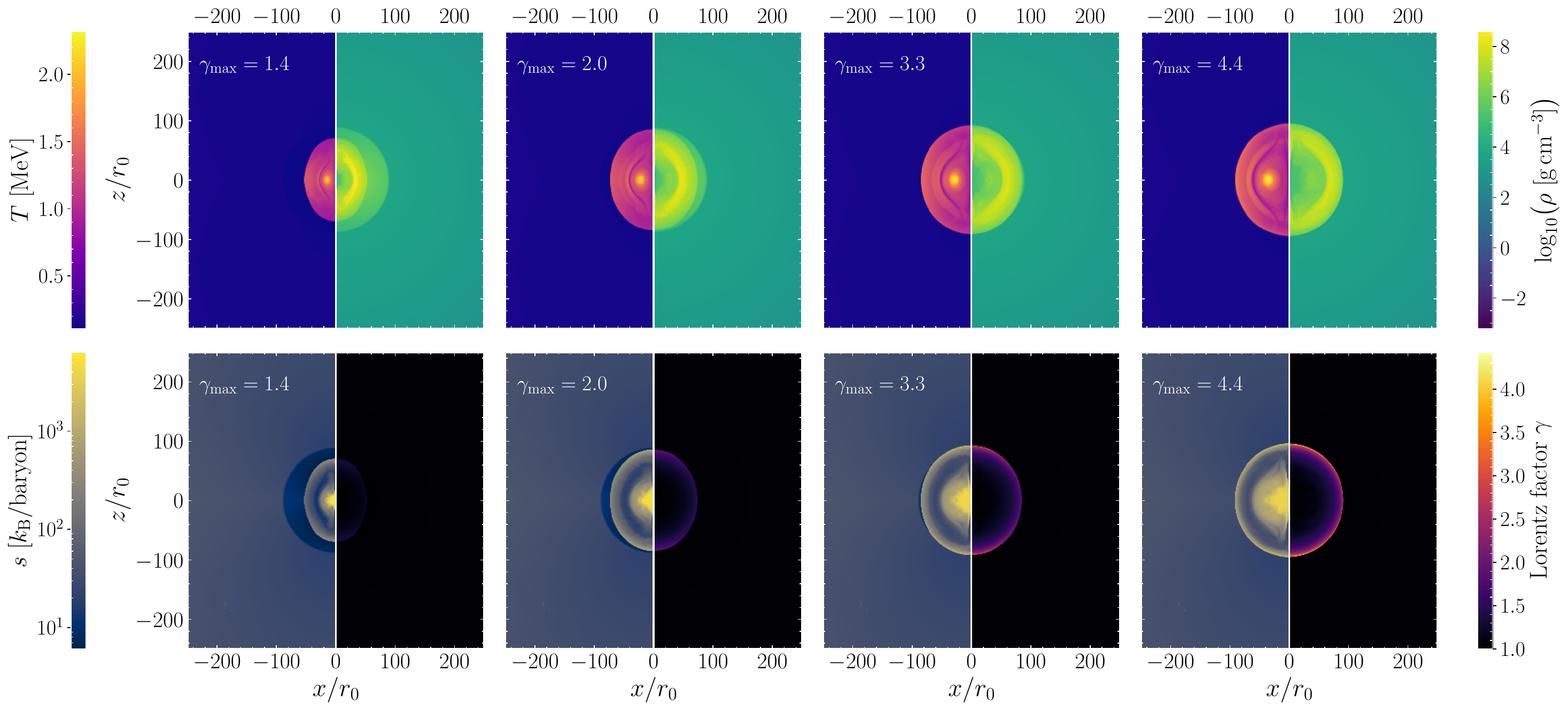}
        \caption*{(a) Early-launch scenario ($t = 101.5\,r_0/c$)}
    \end{subfigure}

    \vspace{0.8em}

    \begin{subfigure}{\textwidth}
        \centering
        \includegraphics[width=\textwidth]{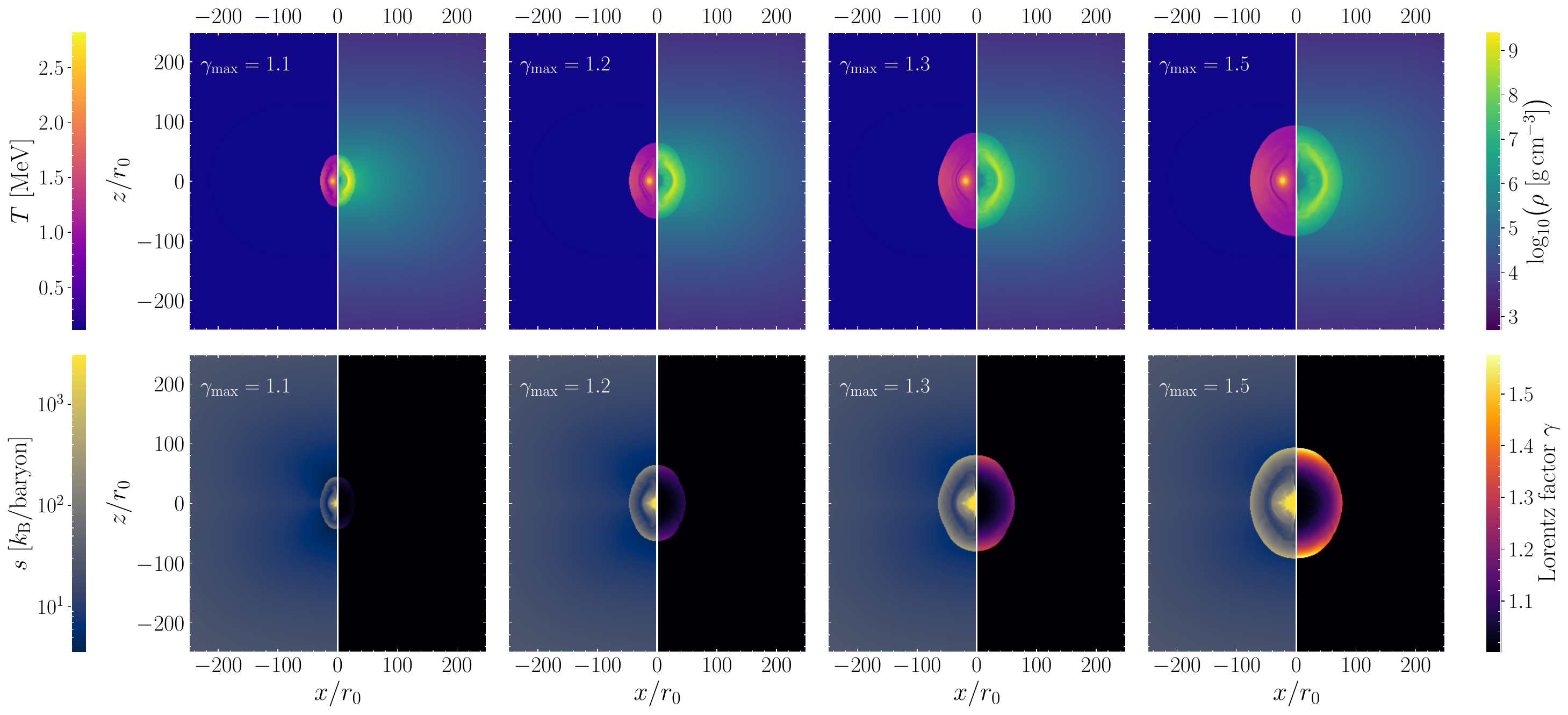}
        \caption*{(b) Late-launch scenario ($t = 121.8\,r_0/c$)}
    \end{subfigure}

    \caption{
    Magnetically powered shock--ejecta interaction for different shock strengths.
    Shown are the final snapshots of the simulations at $ct/r_0 = 101.5$ for the early-launch scenario and $ct/r_0 = 121.8$ for the late-launch scenario, displaying
    the temperature $T$ (left half) and rest-mass density $\rho$ (right half) in the top panels, and
    the entropy $s$ (left half) and Lorentz factor $\gamma$ (right half) in the bottom panels.
    The shock strength is determined by the initial magnetic-field amplitude, which controls the shock acceleration of the ejecta.
    From left to right, the shock strength increases, corresponding to maximum Lorentz factors of $\gamma_{\max} = 1.58$, $2.19$, $2.96$, and $3.90$ for the early-launch case, and $\gamma_{\max} = 1.16$, $1.30$, $1.55$, and $1.84$ for the late-launch case.
    }
    \label{fig:shock_strengths}
\end{figure*}

\section{Results}\label{sec:results}

During and after merger, mass is ejected on multiple timescales.
Early on, {up to about $t\sim20$~ms,} shock heating and tidal forces drive a fast component \citep{Radice:2020ddv, Baiotti:2016qnr}, typically with velocities up to $v\lesssim0.7c$, with the exact values depending mildly on the equation of state \citep{Radice:2020ddv, Lasky:2013yaa, 2024AnP...53600306R}.
On longer timescales ($\sim 200$~ms), the system develops secular ejecta driven by a combination of magnetic \cite{Ciolfi:2017uak, Mosta:2020hlh, Curtis:2023zfo} and neutrino-driven \cite{Metzger:2019zeh, Radice:2020ddv, Fujibayashi:2022ftg, Kawaguchi:2023zln} outflow components, which can arise from the hypermassive neutron star remnant and the accretion disk \citep{Lippuner:2017bfm, Fernandez:2018kax, 2019PhRvD.100b3008M, Curtis:2021guz}.
This latter component is expected to be slower ($v < 0.3c$) but more massive.

Against this background, we launch our initial blast wave and study its interaction with the ejecta.
We consider two launch times that bracket early and late post-merger environments: an ``early-launch'' case, in which the shock is initiated at $t\simeq 40~\mathrm{ms}$ after merger when the ejecta is dense and compact, and a ``late-launch'' case, in which the shock is initiated at $t\simeq 540~\mathrm{ms}$ after merger, following substantial expansion. For each launch time, we perform a parameter study over four shock strengths by varying the initial magnetic field amplitude, and we use the resulting maximum Lorentz factor $\gamma_{\max}$ as the proxy label for shock strength.

Before discussing the shock interactions (Sec. \ref{sec:shock_properties}), the subsequent nucleosynthetic yields (Sec. \ref{sec:nucleosynth}), and the associated kilonova light curves (Sec. \ref{sec:kilonova}), we first summarize the launch of the initial blast wave. In Fig. \ref{fig:sigma}, we show the initial evolution of the blast wave, modeled as an expanding massive plasmoid similar to those ejected in flaring scenarios \citep{Most:2023sft}. The blast wave is initially highly magnetized, but as $\sigma = b^2/(4 \pi \rho c^2)$ drops, it becomes largely hydrodynamic when interacting with the ejecta.

\subsection{Shock properties}\label{sec:shock_properties}

We now provide a detailed assessment of the shock strengths and the resulting reheating of the ejecta cloud.
In Fig. \ref{fig:shock_strengths}, we present a summary of all shock profiles at the time when they have propagated into the ejecta cloud.
The snapshots at $ct/r_0\simeq 100$ (early-launch) and $ct/r_0 \simeq120$ (late-launch) are chosen to coincide with the post-shock phase, i.e., after the shock has traversed the ejecta and the material has cooled to temperatures $T \lesssim 1~\mathrm{MeV}$.
We can see that different shock strengths lead to different anisotropic heating morphologies. In general, weaker shocks are slowed down more in the equatorial direction by the cloud, leading to a prolate profile. The density evolution further illustrates how the ejecta cloud reshapes the blast wave. Weaker shocks are more strongly impeded in the equatorial direction and therefore develop a distinctly prolate post-shock structure, while stronger shocks remain less distorted and evacuate a broader low-density region behind the front. Correspondingly, the Lorentz factor increases systematically with shock strength: the strongest models sustain relativistic expansion over a much larger solid angle, whereas the weaker shocks are confined to a narrower polar channel. The temperature and entropy maps show that this kinematic trend is accompanied by substantial thermodynamic reprocessing. Although the snapshots are taken after the main shock passage, when much of the material has already cooled below $\sim\!1\,\mathrm{MeV}$, the imprint of the reheating remains visible in the extent of the hot and high-entropy layers. Stronger shocks heat a larger fraction of the ejecta to higher temperature and leave behind a shell of high-entropy material, while weaker shocks produce more localized heating and a smaller entropy jump. Since the entropy increase directly traces dissipative shock heating, it provides the clearest measure of how efficiently the blast wave reprocesses the cloud. In this sense, the early-launch models exhibit more global reheating, whereas the late-launch sequence remains more compact and angularly structured, even though the same monotonic trend with shock strength persists.

Having discussed the overall morphology, we now turn to a detailed characterization of the shock impact on the ejecta fluid elements traversed by the shock.
We do this in a Lagrangian framework by following individual fluid elements across the shock. For details on these tracer methods, see, e.g., \cite{Bovard:2017dfh}.



\begin{figure}
    \centering
\includegraphics[width=\columnwidth]{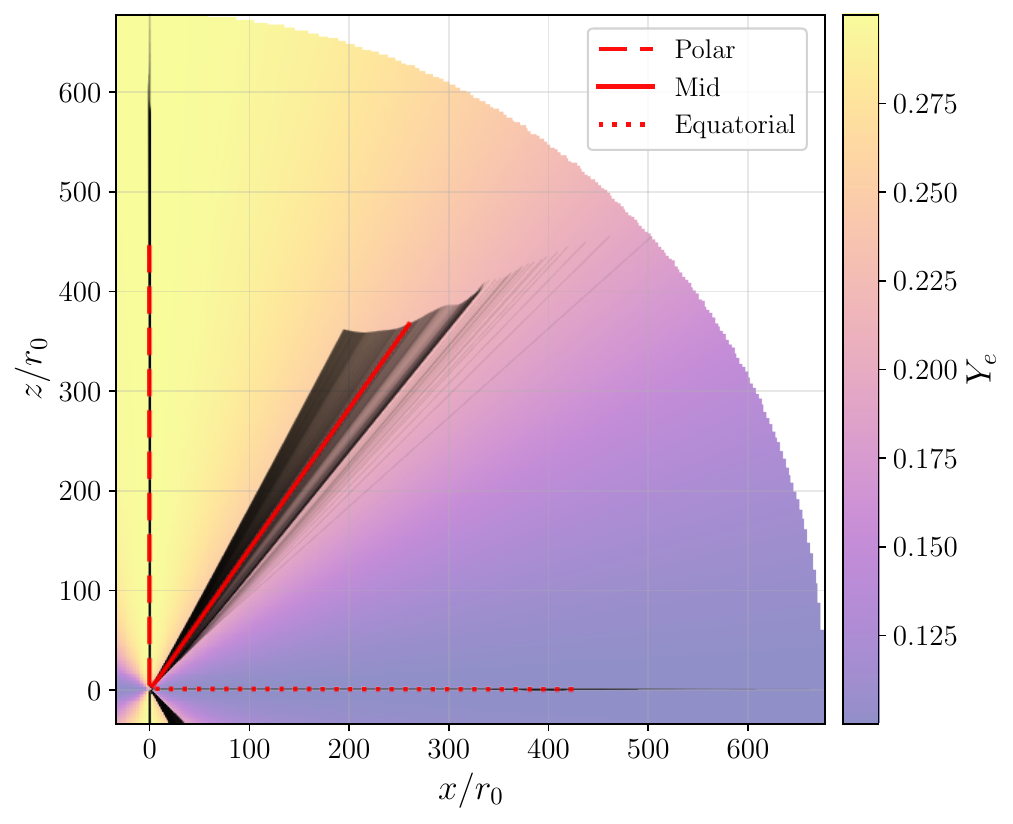}
    \caption{
    Tracer trajectories sampling the ejecta outflow. Shown in gray is the full ensemble of passively advected tracer particles, with 500 tracers sampled in each angular group. The background color indicates the electron fraction $Y_e$. Red lines denote three representative trajectories from the polar, mid-latitude, and equatorial components, distinguished by line style. These trajectories serve as fiducial examples for the thermodynamic and compositional evolution discussed in the following sections.
    }

    \label{fig:tracer_repre}
\end{figure}

We sample fluid tracers along five polar directions ($ \theta = 0^\circ$, $45^\circ$, $90^\circ$, $135^\circ$, $180^\circ$) as outlined in Figure~\ref{fig:tracer_repre}.
Since we reuse these tracers to compute nucleosynthetic yields (Sec. \ref{sec:nucleosynth}), care is required to obtain a representative sample. In each direction, tracers are sampled based on the relative mass weight of the fluid element. For analyzing the hydrodynamic evolution, we use 500 tracers per direction (2500 in total), while for the nucleosynthesis calculations we employ a reduced set of 20 tracers per direction (100 total).
We additionally highlight three fiducial tracers from these groups in Fig. \ref{fig:tracer_repre}, which we use below to illustrate characteristic post-merger evolution. These highlighted tracers span the range of electron fractions in our ejecta, with a relatively high-$Y_e$ polar trajectory, an intermediate-$Y_e$ mid-latitude trajectory, and a low-$Y_e$ equatorial trajectory, providing a compact comparison of composition evolution across angular regions.

\begin{figure*}[t]
    \centering
    \includegraphics[width=\textwidth]{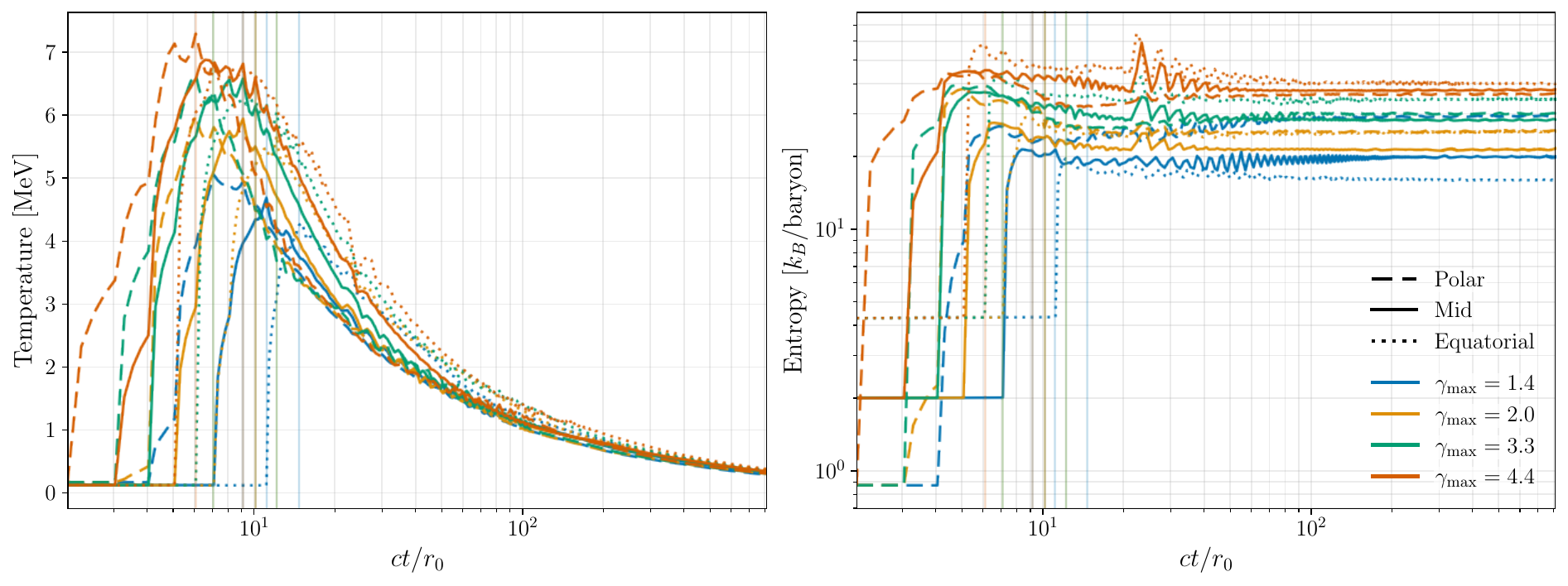}\\[-0.2em]
    {\small (a) Early-launch scenario.}

    \vspace{0.8em}

    \includegraphics[width=\textwidth]{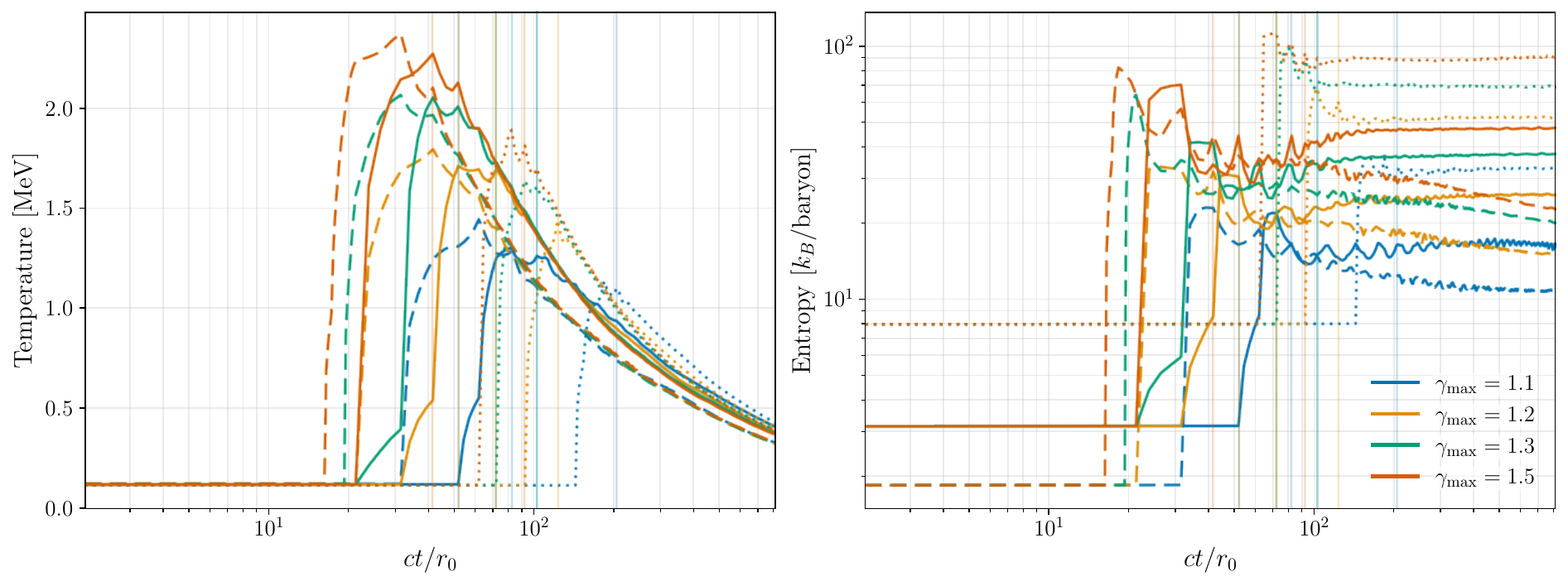}\\[-0.2em]
    {\small (b) Late-launch scenario.}

    \caption{
    Thermodynamic evolution along representative tracer trajectories for the four shock models as functions of time, $t$.
    Shown are ({\it left}) the temperature $T$ and ({\it right}) the entropy per baryon $s$.
    Different line styles encode angular groups (dashed: polar; solid: mid-latitude; dotted: equatorial), and different colors denote the shock strength labeled by the maximum Lorentz factor $\gamma_{\max}$.
    Vertical lines mark the time of shock passage for each tracer, coincident with a rapid rise in $T$ and a step-like increase in $s$, followed by gradual cooling and near-adiabatic expansion at later times.
    }
    \label{fig:selected_tracers}
\end{figure*}

\begin{figure*}[t]
    \centering
    \includegraphics[width=\textwidth]{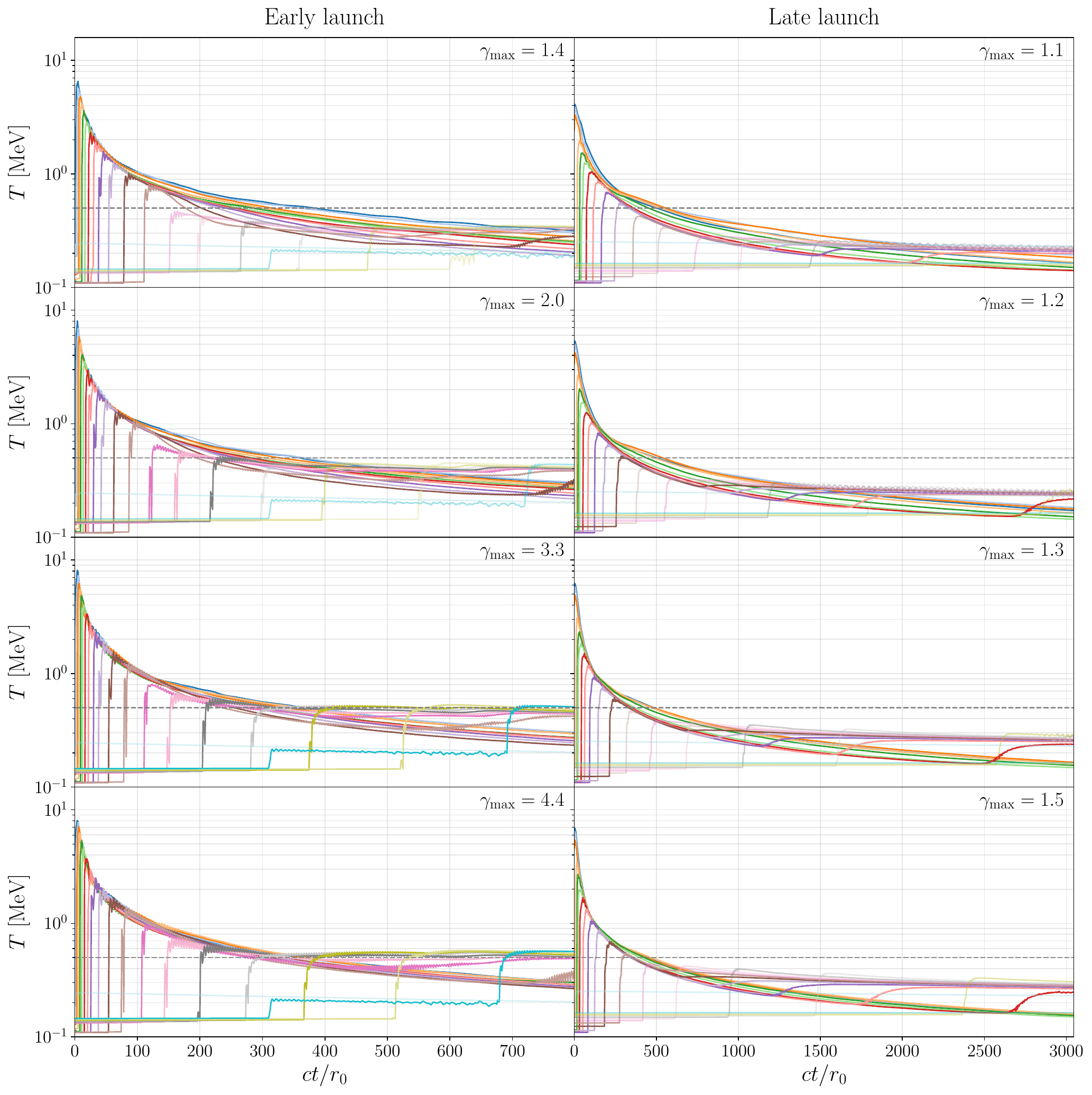}
    \caption{Temperature, $T$, evolution for the entire ejecta profile. Panels are arranged as a $4\times2$ grid: the \emph{left} column shows the \textit{early-launch} series, while the \emph{right} column shows the \textit{late-launch} series. Rows correspond to increasing shock strength, labeled by the maximum Lorentz factor $\gamma_{\max}$ in each panel (early launch: $\gamma_{\max}=1.4,\,2.0,\,3.3,\,4.4$ from top to bottom; late launch: $\gamma_{\max}=1.1,\,1.2,\,1.3,\,1.5$). Within each panel, the colored curves trace individual ejecta trajectories launched from initial radii that are uniformly spaced in $\log r$; as the shock propagates outward, it heats the innermost tracers first and reaches progressively larger radii at later times, as reflected by the staggered temperature rises in the figure. The horizontal dashed line marks $T=0.5~\mathrm{MeV}$; material with $T$ above this value is assumed to be able to reach nuclear statistical equilibrium (NSE); tracers that never exceed this threshold are plotted semi-transparently (faint curves), while tracers that reach $T\ge 0.5~\mathrm{MeV}$ at least once are shown with full opacity. Times $t$ are stated relative to the inner cloud boundary crossing time, $r_0/c$.}
    \label{fig:temp_r_relation}
\end{figure*}

\begin{figure*}
    \centering
    \begin{subfigure}[t]{\columnwidth}
        \centering
        \includegraphics[width=\columnwidth]{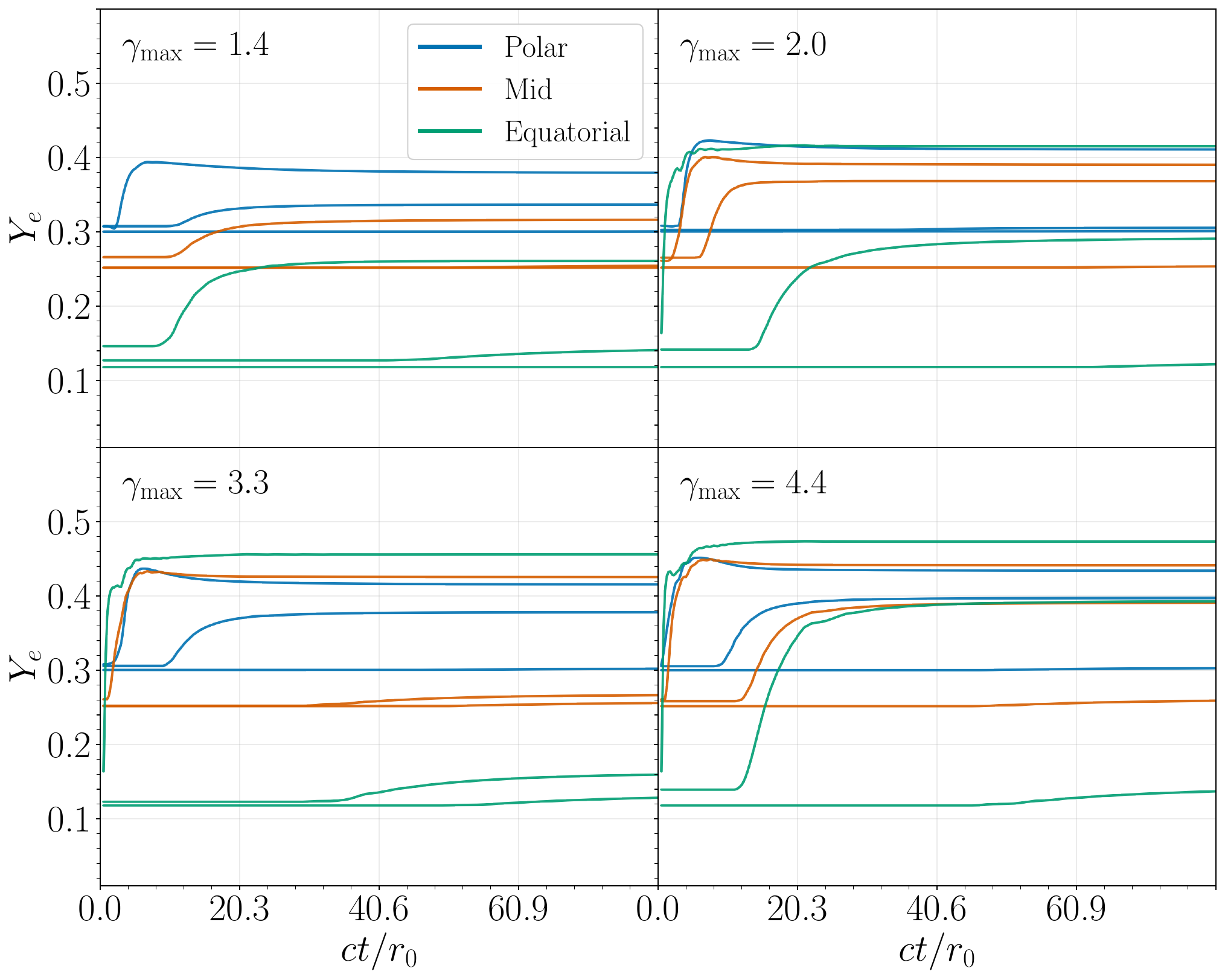}
        \caption{Early-launch scenario.}
        \label{fig:ye_shock_early}
    \end{subfigure}
    \begin{subfigure}[t]{\columnwidth}
        \centering
        \includegraphics[width=\columnwidth]{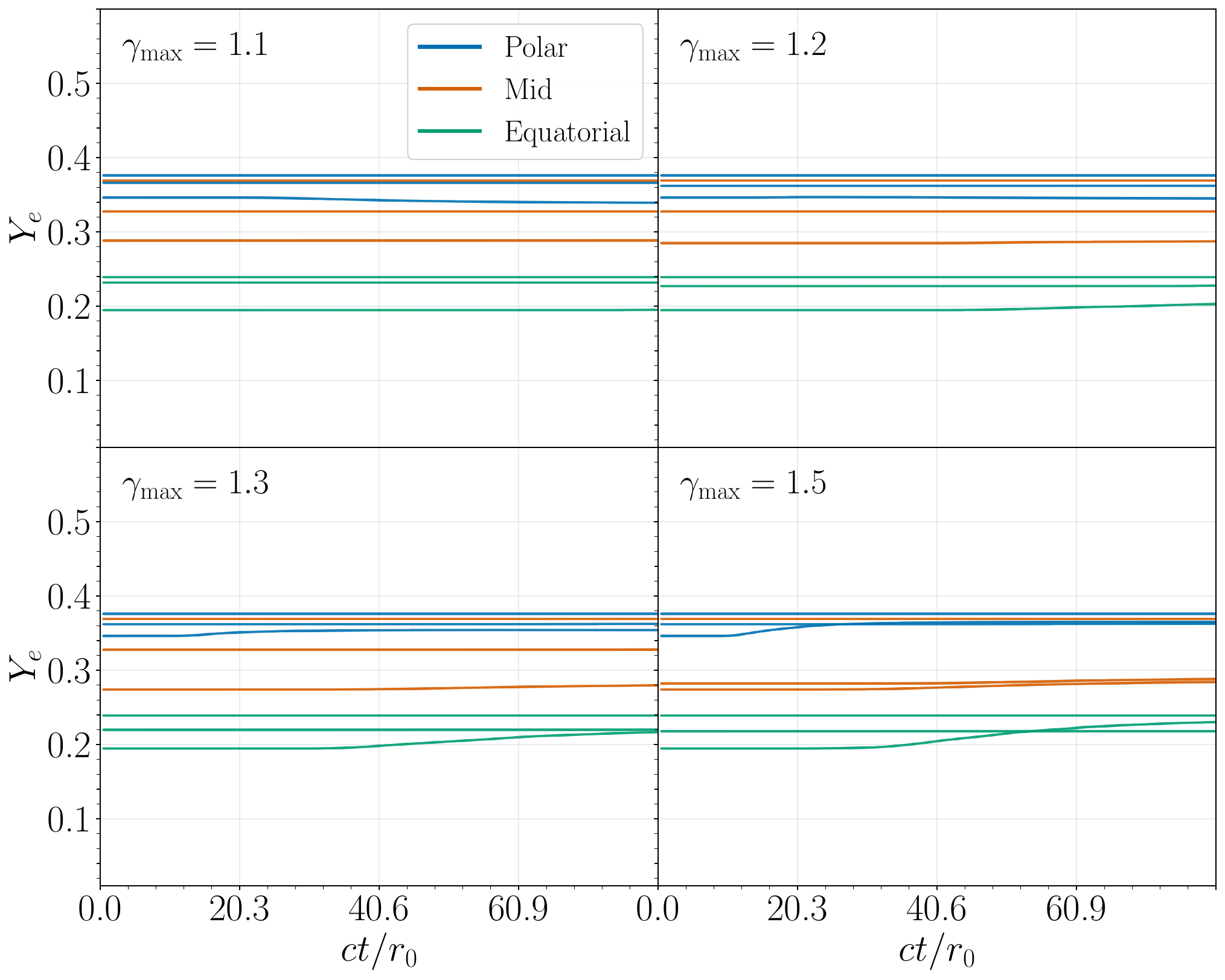}
        \caption{Late-launch scenario.}
        \label{fig:ye_shock_late}
    \end{subfigure}
    \caption{
    Electron-fraction evolution, $Y_e$, for representative tracer particles subject to shock heating. Panels are organized by shock strength, parameterized through the maximum Lorentz factor $\gamma_{\max}$, and colors indicate polar, mid-latitude, and equatorial trajectories. In the early-launch configuration, the shock propagates through denser inner ejecta and heats the material sufficiently for weak interactions to become active. As a result, positron captures efficiently raise $Y_e$ during shock passage, with the magnitude of the increase growing systematically toward stronger shocks (more heating). By contrast, in the late-launch configuration, the ejecta's nuclear composition has already evolved toward higher $Y_e$ due to ongoing nuclear reactions, so that in combination with the weak shocks we consider, the effect of positron capture is small.
    }

    \label{fig:ye_shock}
\end{figure*}

We can now use these tracers to directly study shock heating. Figure~\ref{fig:selected_tracers} summarizes the thermodynamic histories of the three fiducial tracer particles for the four shock models, showing the temperature $T$ and entropy per baryon $s$ as functions of time for the early-launch and late-launch ejecta. In (SR)MHD, the tracers expand homologously along isentropes, so the temperature quickly drops below nuclear statistical equilibrium (NSE) at around $T=0.5\,\rm MeV$. In the absence of additional heating from nucleosynthesis \cite{Metzger:2019zeh}, the ejecta remain ``cold'' before shock interaction, as seen in Fig. \ref{fig:selected_tracers}.

The interaction with the shock causes a sharp temperature jump, sometimes up to $7\, \rm MeV$ in the early-launch models and $3\,\rm MeV$ in the late-launch models, although the details depend most strongly on the initial shock strength. We find that the shock produces specific entropies $10<s/k_B<50$, depending on the model. After shock passage, the ejecta continue to expand along a different isentrope, leading to rapid cooling below the NSE temperature within $10$--$20\, \rm ms$.
Differences between angular groups primarily reflect the ordering of shock arrival and the resulting dispersion in post-shock entropy.
We clarify the latter point in Fig. \ref{fig:temp_r_relation} by considering shock passage through fluid elements at different initial radii, parameterized by the dimensionless quantity $ct/r_0$. Since $r_0/c$ sets the characteristic expansion timescale, the elapsed time in units of the initial light-crossing time provides a natural dimensionless measure of the evolution. We find that weaker shocks decelerate more quickly, so they can only efficiently process the inner ejecta core and become too weak as they approach the outer layers. This means that reheating to NSE---which, as we show below, is the main requirement for affecting the kilonova---operates efficiently only in the inner ejecta zones.
Tracers originating at larger radii remain below this temperature at all times, implying that a substantial fraction of the outer ejecta is not shock-heated enough to enter NSE. The extent of the NSE-processed region grows systematically with shock strength: stronger shocks (higher $\gamma_{\max}$) drive higher peak temperatures and raise a larger subset of inner trajectories above $0.5~\mathrm{MeV}$, directly linking the thermal processing of the ejecta to the shock acceleration measured by the Lorentz factor.\\

\begin{figure*}[t]
    \centering
    \includegraphics[width=\textwidth]{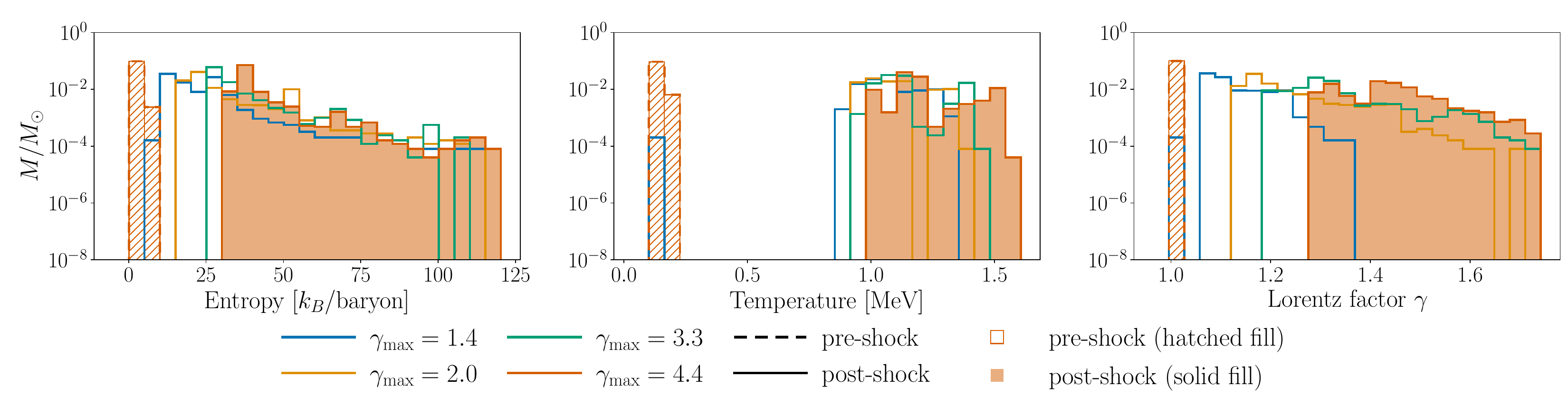}\\[-0.2em]
    {\small (a) Early-launch scenario.}

    \vspace{0.8em}

    \includegraphics[width=\textwidth]{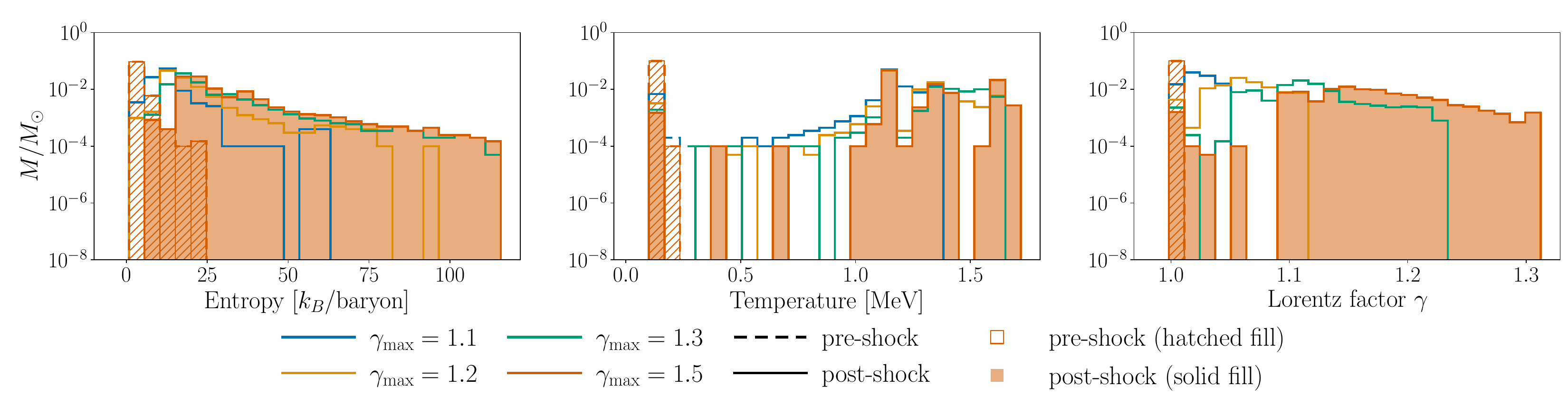}\\[-0.2em]
    {\small (b) Late-launch scenario.}

    \caption{
    Ensemble outflow properties. Shown are ({\it from left to right}) entropy per baryon $s$, temperature $T$, and Lorentz factor $\gamma$. Colors indicate the four shock models labeled by the maximum Lorentz factor $\gamma_{\max}$. For each model, hatched histograms show the pre-shock state, while solid filled histograms show the post-shock state. The shock passage shifts the ejecta to higher entropies and temperatures and accelerates a fraction of the outflow to larger $\gamma$, with stronger shocks producing more pronounced high-$s$, high-$T$, and high-$\gamma$ tails.
    }
    \label{fig:traj_schematic}
\end{figure*}

 One of the main impacts of the shock is to protonize the ejecta
via positron capture \cite{Rosswog:2003rv}
\begin{align}
    n + e^+ \rightarrow p + \bar{\nu}_e\,.
\end{align}
We therefore include this weak interaction in post-processing for the evolution of the electron fraction, $Y_e$. This is done using the rates of Ref. \cite{Rosswog:2003rv} and the implementation in the \texttt{FIL} code \cite{Most:2019kfe, Etienne:2015cea}. Figure~\ref{fig:ye_shock} shows the evolution of $Y_e$ across the shock for representative tracer trajectories in the early-launch and late-launch model sequences. In all cases with sufficient heating, positron capture inside the shocked material drives $Y_e$ upward, but the magnitude of this reprocessing depends sensitively on shock strength (and in turn on tracer position), since a minimum heating level is required to reach NSE. For the density profiles we consider, this is more likely for inner tracers, as the shock rapidly deposits its energy into the ejecta (see Fig. \ref{fig:temp_r_relation}).
For the tracers heated to NSE, $Y_e$ rises rapidly and then saturates at a higher post-shock value (Fig. \ref{fig:ye_shock}). The size of this increase grows with shock strength and is largest for the most neutron-rich equatorial material. In the strongest early-launch models, equatorial tracers are driven from $Y_e \sim 0.12$--$0.15$ to $Y_e \sim 0.4$--$0.47$, while polar and mid-latitude trajectories also exhibit clear, though more moderate, increases.

The major compositional difference in the late-launch ejecta is that nucleosynthesis has already begun for these tracers. In particular, ongoing post-NSE evolution, including $\beta$ decays during the $r$-process, shifts the composition toward higher $Y_e$, so the material is less neutron-rich when it encounters the shock. The additional increase in $Y_e$ induced during shock passage is therefore reduced.
Consequently, the late-launch tracers are initialized using $Y_e$ values from a full nucleosynthesis calculation at the corresponding time (see Section \ref{sec:nucleosynth} for details).
As a result, $Y_e > 0.2$ throughout, and the impact of positron capture is, by construction, smaller.
In turn, the late-launch sequence shows a qualitatively weaker response. For $\gamma_{\max}\lesssim 1.3$, $Y_e$ remains nearly unchanged, with only small secular drifts in selected trajectories. Even for the strongest late-launch model, the overall increase remains modest, and the most neutron-rich equatorial tracers reach only $Y_e \sim 0.23$--$0.24$. This weaker response reflects the fact that, by the time the shock is launched in the late-launch models, the ejecta has already evolved beyond NSE and undergone substantial weak reprocessing.

The comparison between the two sequences shows that the nucleosynthetic impact of the shock depends not only on its strength, but also on the state of the background ejecta. In the early-launch models, the shock substantially raises $Y_e$, especially in the neutron-rich equatorial outflow, so that both weak reprocessing and entropy deposition can alter the final abundance pattern. In the late-launch models, by contrast, $Y_e$ remains largely unchanged, and the shock affects the composition primarily through its entropy input.

To better understand the overall impact of the shock, we now look at ensemble properties of all fluid tracer elements extracted from our simulations (Figure~\ref{fig:traj_schematic}).
We specifically compare pre- and post-shock properties to quantify the impact of shock heating on ejecta evolution.
The resulting ensemble tracer distributions in entropy per baryon $s$, temperature $T$, and Lorentz factor $\gamma$ show that, across all four shock models, shock passage systematically shifts the ejecta to higher entropies and temperatures while accelerating a subset of the outflow to larger $\gamma$. The magnitude of this processing increases with shock strength: stronger shocks generate broader post-shock distributions, more extended high-$s$ and high-$T$ tails reaching $s \gtrsim 10^2\,k_B\,\mathrm{baryon}^{-1}$ and $T \gtrsim 1~\mathrm{MeV}$, and progressively more prominent high-$\gamma$ components, albeit at relatively low Lorentz factors ($\gamma \ll 2$) even for strong shocks. While the temperature almost always jumps to around $1\,\rm MeV$, indicating reheating, the entropy profile remains broad and continuous.
The shock strength therefore controls not only the characteristic post-shock thermodynamic state, but also the fraction of ejecta processed into a high-entropy, high-temperature, and relativistically accelerated component.

\begin{figure*}[t]
    \centering
    \includegraphics[height=0.55\textheight,keepaspectratio]{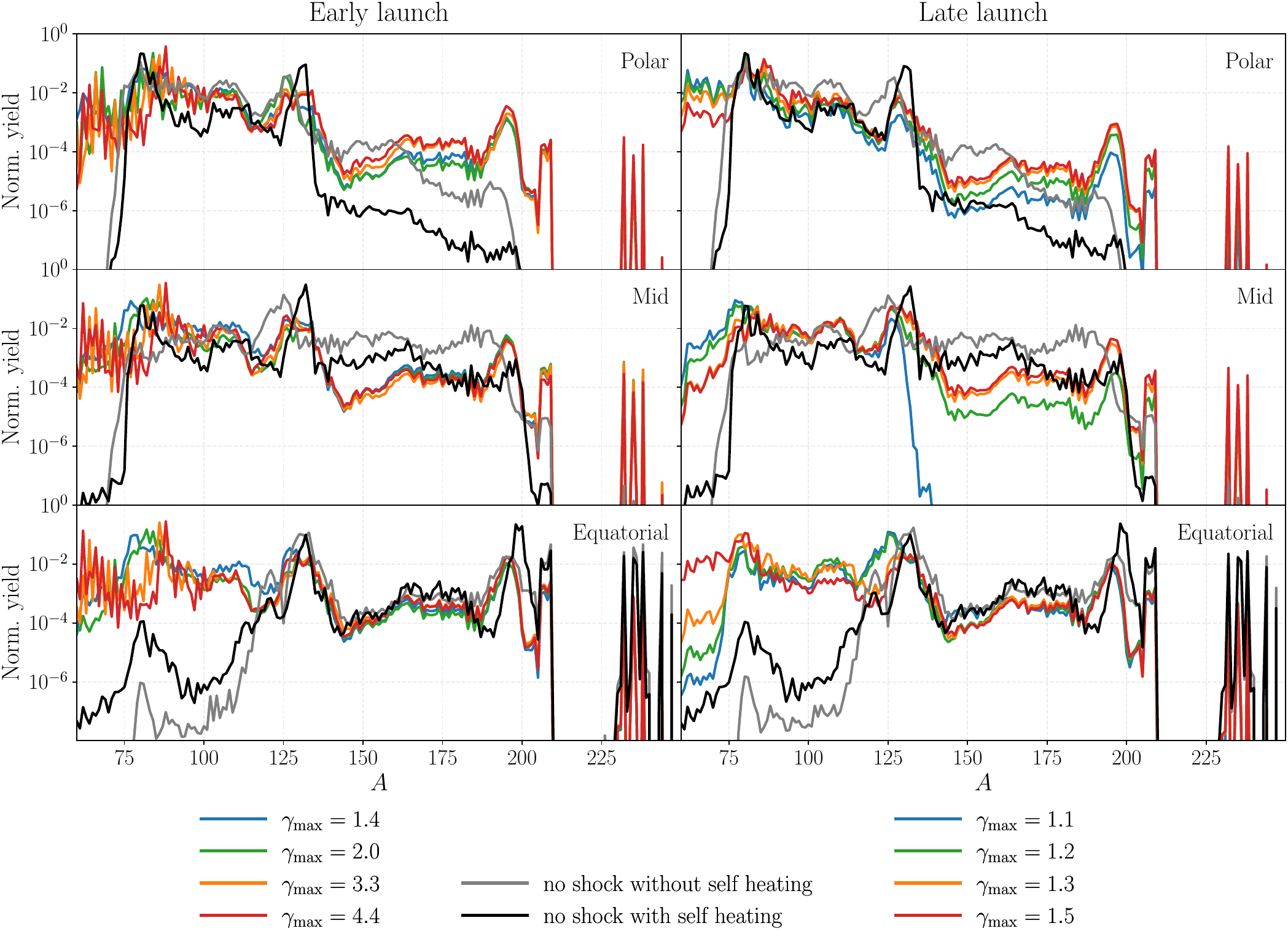}

    \caption{Nucleosynthesis yield patterns for three angular tracer groups (rows: polar, mid-latitude, and equatorial). The left column shows the final normalized yields as a function of atomic mass number $A$ for the early-launch scenario, while the right column shows the corresponding late-launch scenario. Colored curves correspond to shock runs labeled by the maximum Lorentz factor $\gamma_{\max}$ (legend), where $\gamma_{\max}$ is measured at the end snapshot of the underlying MHD simulation for each model. Gray and black curves denote the no-shock reference runs with two different heating prescriptions, without and with self-heating, respectively.}
    \label{fig:yield_compare}
\end{figure*}




\begin{figure*}
\centering
\includegraphics[width=0.9\textwidth]{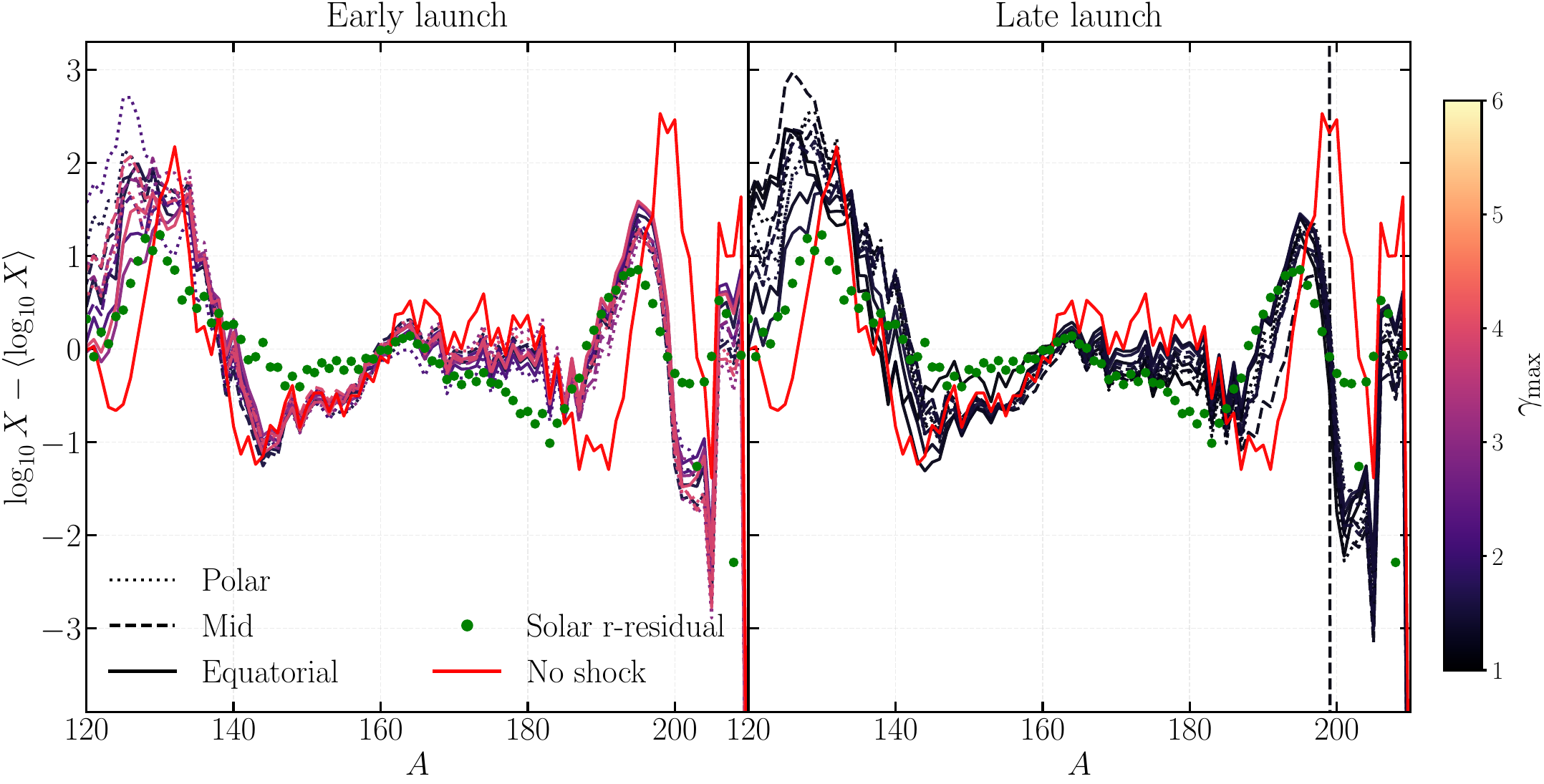}
\caption{
Normalized heavy-element abundance patterns. Shown is the mass-fraction distribution as a function of mass number $A$, normalized according to $\log_{10} X - \langle \log_{10} X \rangle$ over $120 \leq A \leq 210$.
Different line styles denote polar, mid-latitude, and equatorial angular bins, while colors indicate the shock strength parameterized by the maximum Lorentz factor $\gamma_{\max}$. Green points show the solar $r$-process residual normalized in the same way, and the red curve shows the no-shock reference model. The left and right panels show the early-launch and late-launch scenarios, respectively. Across both models, the resulting abundance distributions exhibit only modest angular and shock-strength dependence in the heavy-element regime, indicating that the overall shape of the lanthanide and third-peak pattern remains comparatively robust.
}
\label{fig:yield_solar}
\end{figure*}

\begin{figure*}[t]
\centering
\begin{subfigure}{0.9\textwidth}
  \centering
  \includegraphics[width=\linewidth]{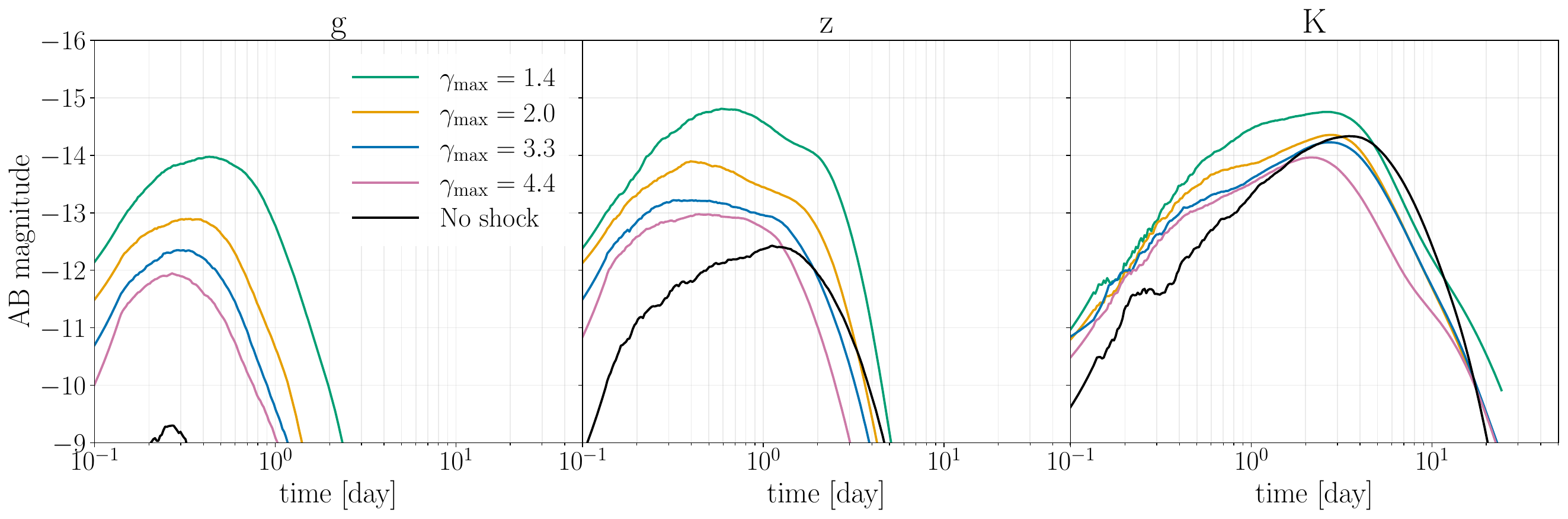}
  \caption{Early-launch scenario.}
\end{subfigure}

\vspace{0.8em}

\begin{subfigure}{0.9\textwidth}
  \centering
  \includegraphics[width=\linewidth]{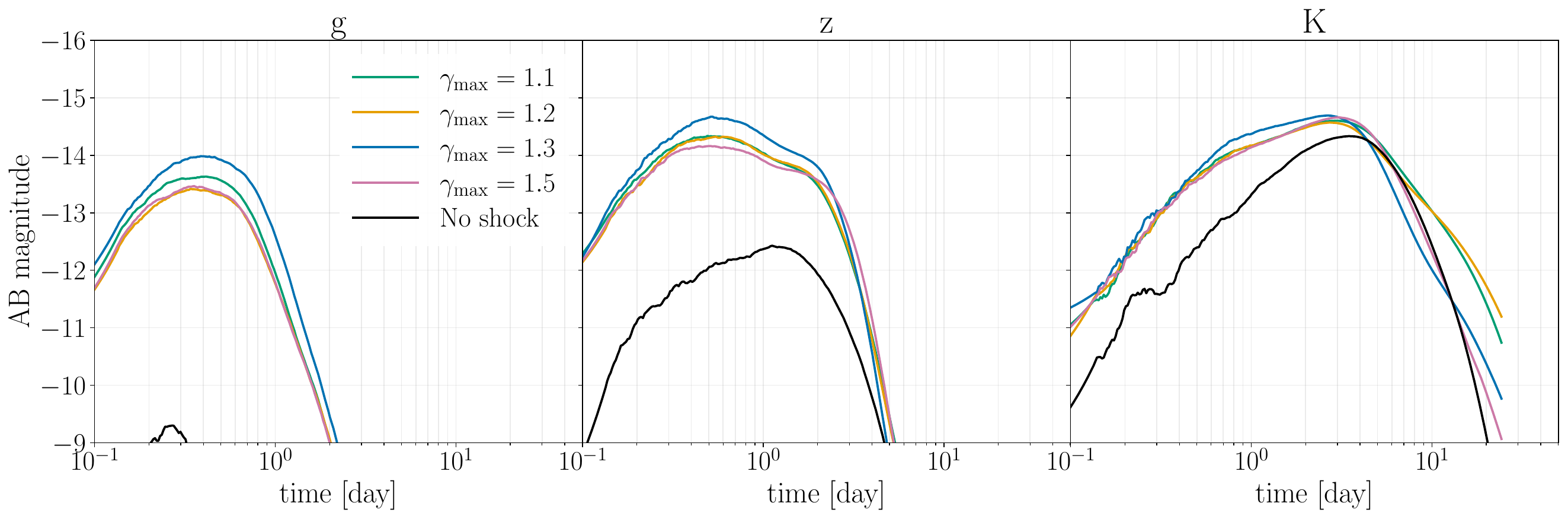}
  \caption{Late-launch scenario.}
\end{subfigure}
\caption{
Synthetic kilonova light curves in the $g$, $z$, and $K$ bands for the early-launch ({\it top}) and late-launch ({\it bottom}) models. Colors denote different shock strengths, parameterized by the maximum Lorentz factor $\gamma_{\max}$, and the black curve shows the no-shock reference model. In all models, the emission shifts from an early, rapidly fading optical component to longer-lived red and near-infrared emission. The shock-processed models are generally brighter than the no-shock reference in the optical bands, while the ordering among shock strengths depends on band and launch scenario.
}
\label{fig:kilonova_band}
\end{figure*}

\begin{figure}[!t]
    \centering
    \begin{subfigure}[t]{\columnwidth}
        \centering
        \includegraphics[width=\columnwidth]{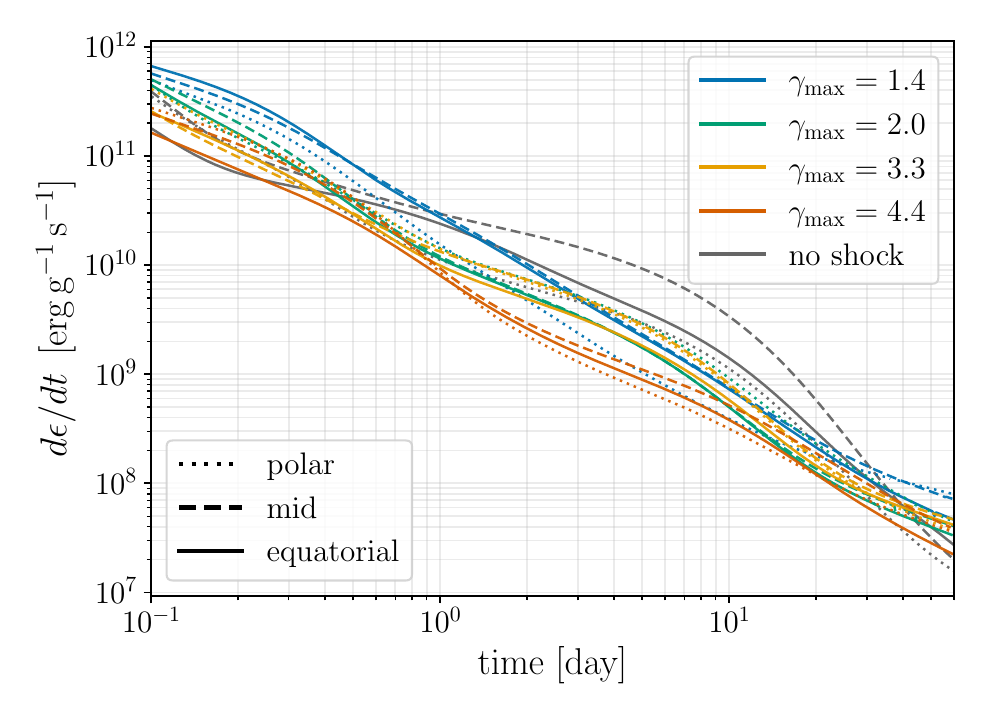}
        \caption{Early-launch scenario.}
        \label{fig:generated_energy_early}
    \end{subfigure}
    \vspace{0.2em}
    \begin{subfigure}[t]{\columnwidth}
        \centering
        \includegraphics[width=\columnwidth]{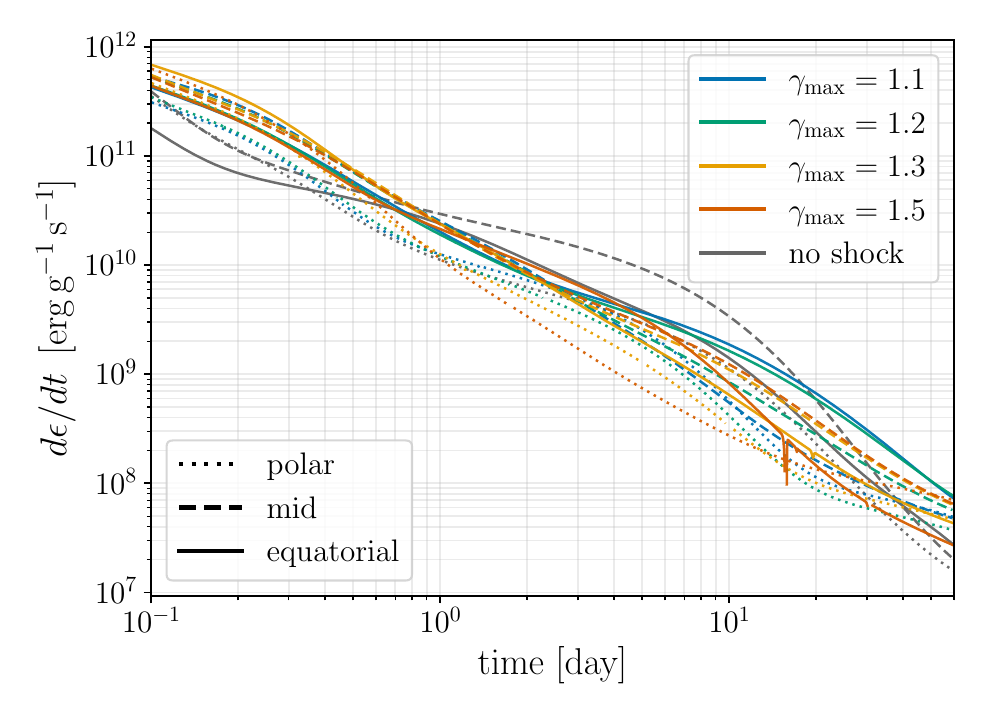}
        \caption{Late-launch scenario.}
        \label{fig:generated_energy_late}
    \end{subfigure}
    \caption{Time-dependent radioactive heating rates, $d\epsilon/dt$, during nucleosynthesis for different angular outflows. Colors indicate different shock strengths, parameterized by the maximum Lorentz factor $\gamma_{\max}$, while line styles denote polar, mid-latitude, and equatorial components. Gray curves show the corresponding no-shock reference model. The upper and lower panels display the early-launch and late-launch cases.
    }
    \label{fig:generated_energy}
\end{figure}

\subsection{Nucleosynthesis}\label{sec:nucleosynth}

 An important aspect of mass ejection from neutron star mergers is the synthesis of heavy elements via the r-process \cite{Eichler:1989ve, Li:1998bw, Metzger:2019zeh}.
 Having established that our different shock scenarios can critically alter the nuclear composition of the ejecta, it is important to propagate this change to the nuclear abundance pattern resulting from the merger and, in turn, to potentially observable changes in the kilonova light curve (see Section \ref{sec:kilonova}).

 A large body of literature has been devoted to calculating nucleosynthesis from merger outflow simulations, incl. \cite{Metzger:2019zeh, Radice:2020ddv, Drout:2017ijr, Lippuner:2017bfm, Curtis:2021guz, Fujibayashi:2022ftg}.

 In brief, we decouple the fluid tracers from the simulation after shock heating and continue to evolve them under the assumption of homologous expansion.
 For the abundance calculations, the trajectories are extended to $t=1\,\mathrm{Gyr}$ in order to obtain the asymptotic composition, whereas for the light-curve calculations (Section \ref{sec:kilonova}) they are followed to $61\,\mathrm{days}$.


The nucleosynthesis calculation is carried out in several stages. We first evolve the reaction network along the analytically constructed pre-shock trajectories from merger until the shock-launch time, assuming complete thermalization of the nuclear heating. At the interface between the pre-shock and post-shock evolution, we do not enforce continuity of all thermodynamic variables. Instead, we use the pre-shock network calculation only to update the composition: the final pre-shock value of $Y_e$ is assigned to the corresponding post-shock tracer prior to any additional weak processing. We then account for post-shock weak interactions by applying electron and positron captures, thereby obtaining the updated post-shock $Y_e$ distribution shown in Figure~\ref{fig:ye_shock}.

The post-shock network evolution is initialized only once the temperature along a given trajectory has dropped below $T=1\,\mathrm{MeV}$. This choice is motivated by numerical limitations of the reaction network at higher temperatures. From that point onward, the network is evolved along the post-shock trajectory, including its homologous-expansion extension, to obtain the final abundance distribution.

To isolate the effect of the shock itself, we also construct a no-shock control model. Starting from the same initial MHD snapshot at the shock-launch time, we generate a purely ballistic trajectory set by integrating backward to merger with the same analytic pre-shock prescription and forward to late times under homologous expansion. Using the same pre-shock network evolution to set the composition at $40\,\mathrm{ms}$, this control model provides a direct baseline against which the impact of shock heating and post-shock weak processing on the final abundance pattern can be assessed.

The actual nucleosynthesis calculations are carried out using the nuclear reaction network WinNet \cite{Reichert:2023xqy}. WinNet is a single-zone reaction network capable of following abundance evolution and nucleosynthetic yields over the wide range of conditions relevant for neutron-star merger outflows. For trajectories that reach sufficiently high temperatures, the composition is first assumed to be in nuclear statistical equilibrium, after which the calculation transitions to the full reaction network as the material cools. This procedure allows us to directly connect the shock-modified thermodynamic evolution to the resulting $r$-process abundance pattern. For the unshocked ejecta, we consider two limiting prescriptions for the treatment of radioactive self-heating. In one case, the energy released by radioactive decays is assumed not to be thermalized back into the outflow. In the other, the decay energy is returned to the material, but with $40\%$ assumed to escape in neutrinos. The range spanned by these two assumptions is taken to approximately bracket the physically relevant evolution of the unshocked ejecta. For all shock-processed ejecta models, we neglect radioactive self-heating during network evolution, under the assumption that shock heating provides the dominant contribution to thermal evolution at this stage.

Figure~\ref{fig:yield_compare} shows the final normalized abundance patterns for the three angular tracer groups, presented as functions of the atomic mass number $A$ for the four shock models indicated in color by $\gamma_{\max}$, together with two no-shock reference solutions corresponding to different treatments of radioactive self-heating. In all cases, the resulting distributions remain broadly $r$-process--like, with pronounced structure around the first ($A \simeq 80$), second ($A \simeq 130$), and third ($A \simeq 195$) peaks.

Shocks influence the composition in two competing ways: weak interactions (e.g., positron capture) can increase $Y_e$, reducing neutron richness and suppressing the heaviest $r$-process nuclei; shock heating can also raise entropy, which can favor heavy-element synthesis in sufficiently neutron-rich material. This competition is evident in Figure~\ref{fig:yield_compare}. For the first peak near $A\sim 80$, the enhancement is strongest in the equatorial direction and weaker in the polar direction, closely tracking the magnitude of the shock-induced $Y_e$ increase. In the equatorial direction, the no-shock baseline has $Y_e\sim 0.1$, so first-peak production is strongly suppressed. After shock passage, however, the temperature rise increases the abundance of thermal $e^\pm$ pairs and accelerates charged-current weak reactions; in particular, positron captures on neutrons, $e^{+}+n\rightarrow p+\bar{\nu}_e$, drive the composition to higher $Y_e$, while neutrino absorption on neutrons may also contribute when irradiation is relevant. This boosts production around the first peak. In the mid-latitude and polar directions, where the baseline $Y_e$ is already high enough to produce the first peak, the enhancement is present but weaker.

For the third peak near $A\sim 195$, the directional trend is reversed: enhancement is more pronounced in the polar direction and weakest in the equatorial direction. This reflects the combined role of entropy deposition and $Y_e$ shifts. In the polar and mid-latitude directions, the no-shock baseline $Y_e$ is comparatively high, which disfavors third-peak production. Shock passage deposits substantial entropy, and the resulting increase favors a more $\alpha$-rich freeze-out and suppresses seed formation during the charged-particle phase \cite{Metzger:2019zeh}. By increasing the neutron-to-seed ratio at fixed composition, this process supports heavier-element production near the third peak. In the equatorial direction, both shocked and unshocked cases retain a third peak, but through different channels: the entropy effect partly compensates for the suppression expected from higher post-shock $Y_e$. The resulting equatorial abundance shape also becomes more consistent with the polar and mid-latitude distributions, supporting an entropy-enhanced interpretation rather than a purely low-$Y_e$ channel.

Finally, we compare the resulting nucleosynthetic yields with the solar abundance pattern \cite{Metzger:2019zeh}. While there is still debate over what fraction of heavy elements in the universe can be associated with neutron star mergers alone (e.g., \cite{2019PhRvD.100d3011C, Radice:2020ddv}), our simulations provide an estimate of potential model-specific uncertainties. By comparing the resulting yields (Fig. \ref{fig:yield_solar}), we find that the third peak is slightly shifted to $A<200$ and is considerably reduced relative to the no-shock case. This makes the shock cases overall slightly more comparable to the solar $r$-process abundance pattern, while differences between individual shock strengths remain small, as outlined above. As a result, shock-enhanced models such as those discussed here could primarily be interpreted as a binary switch, under the assumption of sufficient shock strength.

\subsection{Kilonova light curves}\label{sec:kilonova}

We now connect changes in nuclear abundances to potentially observable changes in the associated kilonova transients. Recent spectroscopic abundance-retrieval work further links lanthanide content and inferred element patterns directly to kilonova observables \citep{Vieira:2022tnm, Vieira:2023rbc}.

To this end, we post-process the nucleosynthesis results using radiative-transfer calculations carried out with the code {\sc SuperNu} \cite{Metzger:2019zeh}. The optical, UV, and near-infrared emission is powered by radioactive decay in the freshly synthesized material, while the emergent light curve is controlled by the ejecta expansion and its composition-dependent opacity \cite{Metzger:2019zeh, Barnes:2016umi, Kasen:2017sxr}.

The radiative-transfer model is constructed directly from the nuclear network output. In particular, the radiative-transfer model uses both the time-dependent radioactive heating rates (Fig. \ref{fig:generated_energy}) and the final compositional information, together with tabulated atomic opacities and heating prescriptions \cite{Metzger:2019zeh, Magistrelli:2024zmk, Magistrelli:2025xja, Ma:2025ouj}. {\sc SuperNu} then solves the multigroup Monte Carlo transport problem on the homologously expanding background and returns synthetic kilonova light curves. For each angular direction, we construct representative heating-rate and abundance profiles by density-weighting over all tracer particles evolved in the nucleosynthesis calculation up to 61 days. This defines an effective angular composition and radioactive power source for the radiative-transfer calculation, allowing the imprint of shock processing on the ejecta to be mapped directly to the emergent emission.

The synthetic kilonova light curves obtained from the shock-processed ejecta for the early- and late-launch sequences show the same qualitative trend discussed below. In both cases, the emission retains the characteristic kilonova behavior of a more rapidly evolving optical component and a broader near-infrared component, reflecting the combined effect of radioactive heating and composition-dependent opacity. The key result, however, is that the shock introduces a clear differential imprint between the two launch scenarios. In the early-launch models, the light curves respond strongly to the shock: with increasing $\gamma_{\max}$, the emission shifts toward earlier times and bluer bands, producing systematically brighter optical light curves and a reduced relative prominence of the longer-lived infrared emission. In the late-launch models, by contrast, the light curves cluster much more tightly, remain closer to the no-shock baseline, and show only a comparatively weak dependence on shock strength.
We caution, however, that the range in shock strengths considered is narrower for the late-launch than for the early-launch cases, while the late-launch sequence evolves primarily through entropy changes rather than background $Y_e$ changes.
One important feature common to both is the substantial enhancement of $g$- and $z$-band emission by several orders of magnitude at peak, although the decay remains rapid within at most a few days. The primary modification in the $K$ band is a change in the late-time decay slope on a timescale of about 10 days.

This difference follows directly from the nucleosynthetic impact of the shock, which can be most cleanly understood in terms of the effective net heating rate (Fig. \ref{fig:generated_energy}). In the early-launch sequence, shock passage modifies both the heating history and the final composition of a substantial fraction of the ejecta. In particular, the upward shift in $Y_e$ suppresses the production of the heaviest $r$-process material and correspondingly reduces the lanthanide-rich contribution to the opacity, allowing radiation to diffuse out earlier and at shorter wavelengths. At the same time, the shock deposits additional entropy and reshapes the thermodynamic trajectories, which further alters the radioactive power input inherited from the nuclear network calculation. The net effect is therefore a kilonova that becomes progressively bluer and faster-evolving with increasing shock strength. The late-launch models likewise show clear departures from the no-shock reference, indicating that the delayed shock still reprocesses the ejecta sufficiently to modify both the radioactive heating history and the effective opacity. Compared to the early-launch sequence, however, these changes are less uniform across angular components and wavelength bands, with the clearest differences appearing in the redder bands. Indeed, we find some of the largest deviations in the heating rate either on a timescale of hours, relevant for the $g$ and $z$ bands, or on a timescale of several days, relevant for late-time slopes in the $K$ band.

While indicative, more systematic parameter coverage will be needed to directly connect our models with kilonova analysis and inference pipelines \citep{Villar:2017wcc, Cowperthwaite:2017dyu, Metzger:2019zeh, 2017ApJ...849...12C, 2019PhRvD.100d3011C, Collins:2023btn, Ricigliano:2023svx}.

\section{Conclusions}\label{sec:conclusions}

In this work, we systematically studied how magnetically driven shocks from long-lived neutron-star merger remnants modify ejecta thermodynamics, nucleosynthesis, and kilonova observables \citep{Most:2023sft, Beloborodov:2022pvn, Most:2024qgc, 2025ApJ...982L..54K}. Using two-dimensional SRMHD simulations, we modeled magnetized blast waves interacting with expanding merger ejecta in both early-launch and late-launch scenarios, and then post-processed tracer trajectories using nuclear-reaction-network and radiative-transfer calculations.

Our main results are as follows. First, shock strength and launch time jointly control the degree of ejecta reprocessing: stronger and earlier shocks produce broader, more anisotropic reheating and larger entropy deposition, while later shocks interact with a more dilute background and leave a comparatively weaker global imprint. Second, sufficiently strong shocks can reheat a subset of the ejecta to near nuclear statistical equilibrium and drive composition changes through both thermal processing and weak interactions, primarily upward shifts in $Y_e$ due to positron capture. Third, these thermodynamic and compositional changes propagate to observables, yielding measurable differences in kilonova color evolution, peak brightness, and late-time decay behavior, with the strongest effects found in the early-launch scenario.

Taken together, our results indicate that remnant variability in the form of magnetic flares and collapse-driven shocks can be a source of intrinsic kilonova diversity and a potential systematic uncertainty for kilonova inference pipelines \citep{Villar:2017wcc, Cowperthwaite:2017dyu, Metzger:2019zeh, Collins:2023btn, Ricigliano:2023svx}. A key limitation of the present work is the two-dimensional approximation: while it is computationally efficient for parameter exploration, it cannot capture non-axisymmetric turbulence, 3D mixing, and magnetic structure evolution seen in three-dimensional merger calculations \citep{Kiuchi:2015sga, Ciolfi:2017uak, Mosta:2020hlh, Collins:2022ocl, Shingles:2023kua}.

Future work should therefore couple long-term remnant evolution to ejecta interaction in fully three-dimensional settings, including self-consistent jet/ejecta coupling and broader microphysical uncertainty quantification \citep{Radice:2020ddv, 2024PhRvD.109f4061N, 2025PhRvD.111j3043J, Lippuner:2017bfm, Fernandez:2018kax, Curtis:2021guz, Curtis:2023zfo, Fujibayashi:2022ftg, Kawaguchi:2023zln, Gottlieb:2017pju, Xie:2018vya, Murguia-Berthier:2017kkn, Pais:2024mpw}. Such extensions will be essential for identifying and interpreting potential magnetic shock (or similar energy injection) imprints in future multi-messenger detections of neutron star coalescence.

\begin{acknowledgments}
The authors gratefully acknowledge insightful discussions with A. Beloborodov, D. Brethauer, T. Govreen-Segal, J. Granot, D. Kasen, E. Nakar, and are grateful to R. Wollaeger for detailed comments on the manuscript. YF and ERM acknowledge support from NASA's ATP program under grant 80NSSC24K1229 and from the National Science Foundation under grant No. PHY-2309210. AFS acknowledges support through Caltech's WAVE Fellows program. YF acknowledges support through the Caltech ZTF--LANL summer school program. CJF acknowledges support from Los Alamos National Laboratory, which is operated by Triad National Security, LLC, for the National Nuclear Security Administration of the U.S. Department of Energy under Contract No.~89233218NCA000001. 
This research was supported in part by grant NSF PHY-2309135 to the Kavli Institute for Theoretical Physics (KITP).
Simulations were performed on the NERSC Perlmutter cluster through allocations m4575
and m5081. This research used resources of the National Energy Research Scientific
Computing Center, which is supported by the Office of Science of the U.S.
Department of Energy under Contract No. DE-AC02-05CH11231.
Software used in this work includes {\sc kuibit} \citep{Bozzola:2021hus}, matplotlib \citep{2007CSE.....9...90H}, numpy \citep{2020Natur.585..357H}, and scipy \citep{2020NatMe..17..261V}.
\end{acknowledgments}

\bibliographystyle{apsrev4-2}
\bibliography{inspire,non_inspire}






\end{document}